\documentclass[reprint, amsmath, amssymb, aps, superscriptaddress,pra]{revtex4-2}

\usepackage{hyperref}
\hypersetup{colorlinks=true,linkcolor=blue,citecolor = blue, urlcolor=blue}
\usepackage{graphicx}
\usepackage[utf8]{inputenc}
\usepackage{amssymb}
\usepackage{amsthm}
\usepackage{mathrsfs}
\usepackage{dsfont}
\usepackage{ upgreek }
\usepackage{float}
\usepackage{physics}
\usepackage{cancel}
\usepackage{quantikz}
\usepackage{mdframed}
\usepackage{breakurl}

\usepackage{tikzit}

\tikzstyle{red dot}=[fill=red, draw=none, shape=circle]
\tikzstyle{green dot}=[fill={rgb,255: red,70; green,176; blue,20}, draw=none, shape=circle]
\tikzstyle{Resize}=[font={\scriptsize}]
\tikzstyle{Big}=[font={\huge}]
\tikzstyle{Photon}=[fill={rgb,255: red,255; green,230; blue,103}, draw=black, shape=circle, opacity=0.7]
\tikzstyle{Big node}=[fill={rgb,255: red,191; green,191; blue,191}, draw=black, shape=circle]
\tikzstyle{small node}=[fill={rgb,255: red,191; green,191; blue,191}, draw=black, shape=circle, inner sep=0,minimum size=3pt]

\tikzstyle{Fill red}=[-, fill={rgb,255: red,255; green,184; blue,184}]
\tikzstyle{dashes}=[-, dashed, dash pattern=on 1mm off 1mm]
\tikzstyle{Fill grey}=[-, fill={rgb,255: red,166; green,166; blue,166}]
\tikzstyle{Thick}=[-, thick]
\tikzstyle{Arrow}=[->]
\tikzstyle{red line}=[-, fill=none, draw=red]
\tikzstyle{Blue line}=[-, fill=none, draw=blue]
\tikzstyle{Thick arrow}=[thick, ->]
\tikzstyle{Fill pink}=[-, fill={rgb,255: red,255; green,140; blue,140}]
\tikzstyle{dashed arrow}=[->, dashed, dash pattern=on 1mm off 0.5mm]
\tikzstyle{arrow}=[->, very thick, draw=red]
\tikzstyle{dashes red}=[-, dashed, dash pattern=on 1mm off 1mm, fill=none, draw=red]
\tikzstyle{dashes blue}=[-, dashed, dash pattern=on 1mm off 1mm, fill=none, draw=blue]
\tikzstyle{Thick dashed arrow}=[->, thick, dashed, dash pattern=on 1mm off 0.5mm]

\usepackage{orcidlink}

\date{\today}

\newcommand{\N}{\mathbb{N}}
\newcommand{\R}{\mathbb{R}}

\newcommand{\1}{\mathds{1}}
\renewcommand{\bar}{\overline}
\renewcommand{\P}{\mathbb{P}}
\newcommand{\comment}[1]{}
\newcommand{\changes}[1]{#1}

\newenvironment{preuve}{\begin{proof}[\underline{Proof}]}{\end{proof}}
\newenvironment{Thm}{\begin{theorem}}{\end{theorem} } 
\newenvironment{Lem}{\begin{lemma}}{\end{lemma}}
\newenvironment{Pro}{\begin{property}}{\end{property}}
\newenvironment{Def}{\begin{definition}}{\end{definition}}
\newenvironment{Cor}{\begin{corollary}}{\end{corollary}}

\newtheorem{theorem}{Theorem}

\newtheorem{lemma}{Lemma}
\newtheorem{property}{Proposition}
\newtheorem{corollary}{Corollary}
\theoremstyle{definition}
\newtheorem{definition}{Definition}

\theoremstyle{remark}

\newcommand{\ie}{\textit{i.e. }}

\begin{document}

\title{Noisy certification of continuous variables graph states}

\author{\'Eloi Descamps\orcidlink{0000-0002-6911-452X}}
\email{eloi.descamps@u-paris.fr}
\affiliation{Universitée Paris Cité, Laboratoire Matériaux et Phénomènes Quantiques, 75013 Paris, France}
\author{Damian MARKHAM\orcidlink{0000-0003-3111-7976}}
\affiliation{Sorbonne Université, CNRS, LIP6, F-75005 Paris, France}

\begin{abstract}
Continuous variables (CV) offer a promising platform for the development of various applications, such as quantum communication, computing, and sensing, and CV graph states represent a family of powerful entangled resource states for all these areas. In many of these protocols, a crucial aspect is the certification of the quantum state subsequently used. While numerous protocols exist, most rely on assumptions unrealistic for physical continuous-variable states, such as infinite precision in quadrature measurement or the use of states requiring infinite squeezing.
In this work, we adapt existing protocols to deal with these unavoidable considerations and use them to certify their application for different quantum information tasks. More specifically, we show how CV graph states can be efficiently verified and certified even in a noisy and imperfect setting. We then discuss how our findings impact the usability of states obtained after the protocol for different applications, including quantum teleportation, computing, and sensing.
\end{abstract}

\maketitle


\section{Introduction}
Quantum information theory unlocks advanced capabilities across diverse areas, spanning computing, sensing, and communication. 
The landscape of quantum technology development encompasses numerous platforms, with different advantages and challenges. 
Typically quantum information is phrased in terms of qubits which are two-dimensional quantum systems \cite{nielsen_quantum_2010}. However, for some physical systems, infinite-dimensional description is more natural, particularly in optics. Such systems have motivated an alternative approach, framing information processing in continuous variables (CV) \cite{braunstein2005quantum}. The CV framework is a major contender for future quantum technologies, in computation \cite{lloyd1999quantum,gottesman_encoding_2001}, communication \cite{furusawa1998unconditional} and sensing \cite{degen2017quantum}. Indeed, CV systems are capable of generating entangled states of sizes orders of magnitude higher than in the discrete, qubit case \cite{yoshikawa2016invited}. 

Graph states are a powerful family of entangled resource states across quantum information, described by graphs, where nodes represent quantum systems and edges represent preparation entanglement.
They exist both for discrete \cite{hein_multiparty_2004} and continuous variables \cite{van2007building}. CV graph states, like the discrete case,  are resources for universal fault-tolerant quantum computing \cite{gu_quantum_2009,menicucci_fault-tolerant_2014,asavanant2019generation,larsen2019deterministic}, secret sharing \cite{tyc2002share,van2011implementing,cai2017multimode}, anonymous communication \cite{menicucci2018anonymous}, distributed sensing \cite{guo_distributed_2020}, error correction \cite{van2011implementing} and more. 
The success of these applications, however, relies on the quality of the graph states. This motivates the crucial question of the certification of graph states, and its application to quantum information processing, which is the main topic of this work.

The certification of quantum resource states is an important problem in quantum information which has been much studied \cite{eisert2020quantum,kliesch2021theory}, in particular in the discrete case, for general resource states \cite{pallister2018optimal,zhu_general_2019,huang2020predicting,govcanin2022sample,fawzi2024learning}, and focused on graph states \cite{pappa2012multipartite,hayashi2015verifiable,markham_simple_2018,takeuchi_resource-efficient_2019}. Indeed such results have not only been used to certify useful resources but integrated into protocols to verify correct behavior for many applications including computation \cite{hayashi2015verifiable,takeuchi_resource-efficient_2019,fujii2017verifiable}, secret sharing \cite{markham2015practical,bell2014experimental}, anonymous communication \cite{unnikrishnan2019anonymity,hahn2020anonymous} and distributed sensing \cite{shettell2022private}.
There are some, but fewer results in CV \cite{aolita_reliable_2015,takeuchi_resource-efficient_2019,chabaud_efficient_2021,wu_efficient_2021,liu_efficient_2021}.
In particular in \cite{takeuchi_resource-efficient_2019} a general protocol is put forward for certifying graph states, which, in principle, works both for discrete and continuous variables. However, its application in CV quickly encounters a core difficulty in the theoretical handling of CV - as it stands, the protocol, though formally correct, will never accept any physical source of states. The reason is, as is common in CV, the CV graph states are, strictly speaking, neither physically nor mathematically meaningful quantum states, except as limits to physical states (infinite squeezing), and similarly the measurements are limits of physical measurements (to nonphysical infinite precision). So that any physical source of states can pass the test, the physical constraints of real states and real measurements must be taken into account. 

In this work, we adapt the protocol and proofs of \cite{takeuchi_resource-efficient_2019} to certify CV graph states to deal with physical states and measurements, and its application for certifying quantum information tasks. The main technical ingredient is a detailed treatment of the measurement POVM and its integration at all steps of the proof of security. Carefully detailing these steps allows us to give different levels of statements of security, allowing for parameters to be optimized to different cases, as well as clearer comparisons to existing results. We then explore the application of the certification of graph states to certify several protocols, including CV quantum teleportation, measurement-based quantum computation, and metrology.

The structure of the paper is as follows: section \ref{Sec: Certification protocol} introduces basic notations, CV graph states, and the certification protocol. In section \ref{sec: security of the certification protocol}, we present a list of mathematical results describing the protocol's performance, with detailed proofs provided in the appendices. Finally, section \ref{sec: application} explores the efficiency of a state verified by our protocol for various quantum information protocols.

\section{Certification protocol} \label{Sec: Certification protocol}
\subsection{Basic definitions}
We begin by establishing basic definitions, starting with that of a weighted graph, a generalization of an unweighted graph more commonly used in DV \cite{hein_multiparty_2004}, which is more natural in CV \cite{menicucci_graphical_2011}. A weighted graph is defined as a triple $G=(V,E,\Omega)$ where:
    \begin{itemize}
        \item $V={1,\dots,n}$ denotes the set of vertices.
        \item $E$ is a set of unordered pairs ${i,j}\subset V$ where $i$ and $j$ are distinct vertices. An element $e\in E$ is called an edge.
        \item $\Omega\in \R^{E}$ is a collection of real numbers $\Omega_e$, indexed by the edges of the graph, assigning a weight $\Omega_e$ to each edge.
    \end{itemize}

Any standard graph $G=(V, E)$ can be treated as a weighted graph by selecting a uniform weight $\Omega$ for all edges, commonly set as $\Omega=1$.

Moving on to quantum notations, we denote quadrature operators as $\hat x$ and $\hat p$ with commutation relations $[\hat x_i,\hat p_j]=i\delta_{i,j}$. Additionally, indices on operators and `bra' and `ket' states designate the mode in which they act or reside. In the continuous variables setting, gates are introduced to generalize the Pauli operators. For $1\leq i,j\leq n$ and $s\in \R$, we introduce the `Z gate' $\hat Z_j(s)$ acting on mode $j$, the `X gate' $\hat X_j(t)$ acting on mode $j$ and the `controlled Z' gate $\hat{CZ}_{i,j}(s)$ acting on mode $i$ and $j$, with expressions
\begin{align}
    \hat Z_j(s)=e^{i s \hat x_j} && \hat X_j(t)=e^{i t\hat p_j}\notag
\end{align}
\begin{equation}
    \hat{CZ}_{i,j}(s)=e^{is\hat x_i \hat x_j}.
\end{equation}
 For $1\leq j\leq n$, we also denote by $\ket{0}^p_j$ the (unormalized) eigenstate of $\hat p_j$ with eigenvalue $0$:
\begin{equation}
    \ket{0}^p_j= (x_j\mapsto \frac{1}{\sqrt{2\pi}}).
\end{equation}

We can now define a CV graph state \cite{menicucci_graphical_2011} associated with a weighted graph $G=(V, E,\Omega)$.

\begin{Def}
    For a weighted graph $G=(V,E,\Omega)$, we define the associated CV graph state $\ket{G}$ as:
    \begin{equation}\label{eq: graph state def}
        \ket{G}=\left(\prod_{\{i,j\}\in E} \hat{CZ}_{i,j}(\Omega_{i,j})\right)\bigotimes_{j=1}^n\ket{0}^p_j.
    \end{equation}
\end{Def}
These CV graph states are characterized by their nullifiers, akin to how finite-dimensional graph states are characterized by their stabilizers \cite{menicucci_graphical_2011}.
\begin{Pro}\label{pro: stab cv graph}
    For a given weighted graph $G=(V, E,\Omega)$, the state $\ket{G}$ is the only state (up to a global phase) that is nullified (\ie eigenvector with eigenvalue $0$) by the operators $\hat g_i$ for all $1\leq i\leq n$ defined by:
    \begin{equation}
        \hat g_i=\hat p_i-\sum_{j\in N^{(i)}}\Omega_{i,j} \hat x_j,
    \end{equation}
    where the set $N^{(i)}\subset V$, is the set of the neighbors of $i$: $N^{(i)}=\{j\in V|\{i,j\}\in E\}$. The operators $\hat g_i$ are called the nullifiers of $\ket{G}$; they completely characterize the state.
\end{Pro}
\begin{preuve}
    We still provide proof as the presentation differs from \cite{menicucci_graphical_2011}. This is straightforward algebra computation, but for completeness, we provide a complete derivation in the appendix \ref{proof: stab cv graph}.
\end{preuve}

The nullifier condition, and its discrete version the stabilizer condition, are how graph states are tested in the existing certification protocols \cite{pappa2012multipartite,hayashi2015verifiable,markham_simple_2018}, including, in particular for the CV case in \cite{takeuchi_resource-efficient_2019}. However, the probability that any physical state (such as the finitely squeezed version of the graph state Eq.~(\ref{eq: graph state finite squeezing}) satisfies the nullifier condition exactly, is zero. \changes{To understand this fact, one can first notice that momentum eigenstates are not physical. As such the momentum distribution of any physical state must have some thickness and the probability to measure a momentum of exactly $0$ is zero. As graph states are obtained by the application of entangling gates on the zero momentum eigenstate (see Eq.~\ref{eq: graph state def}), the nullifier distribution behaves in the same way. As such, the probability of measuring a specific value for the nullified is zero.  This crucial} observation highlights the need for careful consideration of physical states and measurements, and in particular must be taken into account in the protocol itself.

To characterize how close a state is to the ideal graph state $\ket{G}$, it is sometimes useful to fix a figure of merit.
This is a non-trivial choice. For various reasons that will become clearer, we opt for the following definition, which we call overlap, which can be thought of as a fidelity with a Gaussian envelope around it.  In particular, as we will see in section~\ref{sec: application}, this figure of merit can be used to apply the protocol to several different scenarios and problems directly. We define the \textit{overlap }
\begin{equation} \label{eq: overlap}
    O^\Delta=\int d^n s e^{-\norm{\va{s}}^2/\Delta^2}\bra{G}\hat Z(\va{s})\hat\rho' \hat Z(-\va{s})\ket{G}.
\end{equation}
for which, state closed to $\ket{G}$  satisfy $O^\Delta\simeq1$. This quantity is parameterized by the quantity $\Delta$ and measures better the distance to the ideal graph state for small values of $\Delta$. \changes{The physical interpretation of $\Delta$ is discussed in more details in the following. However, equations presented in Sec.~\ref{sec: consequence on nullifier measurement} allow the following one: $\Delta$ correspond roughly to the expected thickness in the distribution of the nullifier.} A similar quantity can be defined to measure the distance between a state and $k$ copies of the ideal graph state
\begin{widetext}
    \begin{equation}
        O^\Delta_k=\int d\va s_1 \cdots d\va s_k \exp(-\frac{\norm{\va s_1}^2+\cdots+\norm{\va s_k}^2}{\Delta^2})\bra{G}^{\otimes k}\hat Z(\va s_1,\dots,\va s_k)\hat \rho'\hat Z(-\va s_1,\dots,-\va s_k)\ket{G}^{\otimes k}.
    \end{equation}    
\end{widetext}

We emphasize that many of our results are not phrased using such a fixed figure of merit, intentionally so that they may be more easily applied to different situations and choices.
Moreover, when they are, while we make this choice, motivated largely by the applications, our results can be rephrased easily with different preferred figures of merit.

\subsection{The certification Protocol}

\begin{figure}
    \centering
    \scalebox{0.8}{\input{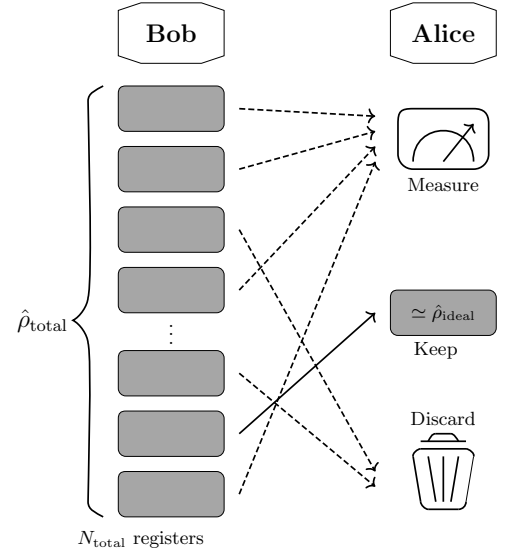}}
    \caption{Generic pictorial representation of certification protocols.}
    \label{fig: generic protocol}
\end{figure}

The problem of certifying a CV graph state can be understood as follows. Alice asks Bob to produce a specific continuous variable graph state, and Alice wants to be sure, for further applications, that the state she received is indeed the one (or at least close enough to the one) she asked for. To achieve this, she employs a state certification protocol. All such protocols essentially follow the same structure \cite{zhu_general_2019}. Alice asks Bob to produce multiple copies (also called registers) of the target state. She will then perform specific measurements on some registers, discard others, and keep one or a few of them at the end. Which register is used for what is decided uniformly at random. After post-processing the measurement, the protocol certifies to Alice whether the states she kept are close enough to the ideal required states, or if the received states are too unreliable. This is illustrated in figure~\ref{fig: generic protocol}. Specific certification protocols are usually tailored to a particular class of states and use the specific characteristics of these states to do the certification. Here, our focus is on certifying CV graph states.

\begin{figure*}
    \centering
    \scalebox{0.72}{\input{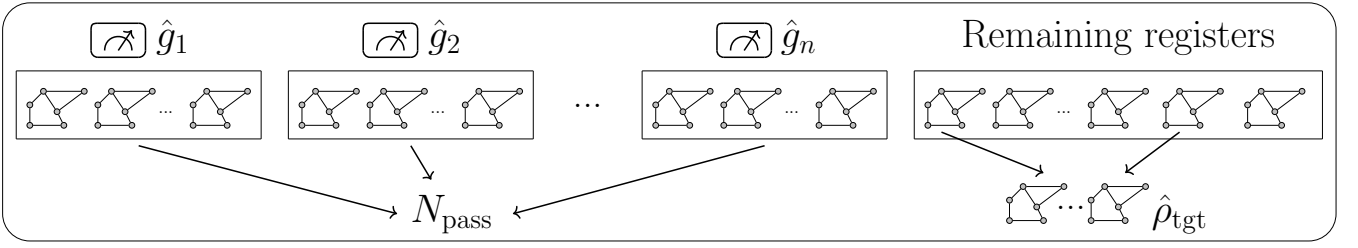}}
    \caption{Schematics of the state certification protocol.   Note that the choice of which copy state is sent to which part of the protocol is random (cf Figure \ref{fig: generic protocol}), which is equivalent to assuming a permutation invariant state and a fixed order, as illustrated.}
    \label{fig: protocol}
\end{figure*}

Let us consider a weighted graph $G=(V, E,\Omega)$ with associated graph state $\ket{G}$. As continuous-variable graph states are defined as operations on the infinitely squeezed momentum eigenstate $\ket{0}^p$, which are non-physical states, Bob can produce only an approximation of these states. We will first consider that Bob can produce
\begin{align}\label{eq: graph state finite squeezing}
    \ket{G^\sigma}&=\frac{1}{(\pi\sigma^2)^{n/4}}\left(\prod_{e\in E} \hat{CZ}_e(\Omega_e)\right)\notag\\
    &~~~~~~\int d^nx e^{-\norm{\va{x}}^2/2\sigma^2}\ket{x_1,\dots,x_n},
\end{align}
where $\sigma$ is the squeezing parameter, with the ideal case corresponding to the limit $\sigma\to\infty$. \changes{Such Gaussian approximation of CV graph states can be produced in the lab with high scalability in the number of modes \cite{larsen2019deterministic}. Furthermore, their usability for different applications such as measurement-based quantum computing has already been studied \cite{gu_quantum_2009}.} The protocol that we propose directly builds on the one of \cite{takeuchi_resource-efficient_2019}. 
The main difference is that we take into account the noisy measurement and the impossibility of a perfect state in the CV setting, which is essential for the protocol to work. Doing so requires new elements to the proof, in particular the description of the accepted POVM, and a small adaptation of the accepting step in the protocol itself to allow for acceptance under physical measurements and states. The protocol runs as follows, and is schematically illustrated in Fig.~\ref{fig: protocol}:\\

\begin{enumerate}
    \item Alice requests a total number of $N_\text{total}$ registers from Bob, each composed of $n$ modes. If Bob cooperates (and if transmission noise is negligible), Alice receives the state $\ketbra{G^\sigma}^{\otimes N_\text{total}}$. However, without assuming Bob's honest behavior, Alice will receive a global state $\hat\rho$, potentially entangled across the registers and even with some ancilla registers stored by Bob.
    
    \item For $i=1,\dots,n$ Alice successively selects $N_\text{test}$ registers randomly (uniformly and without replacement) to measure the nullifier $\hat g_i$. She performs a total of $nN_\text{test}$ measurements. To measure each nullifier $\hat g_i=\hat p_i-\sum_{j\in N^{(i)}}\Omega_{\{i,j\}} \hat x_j$, she simply measure the quadrature operators $\hat p_i$ and $\hat x_j$ ($j\in N^{(i)}$) on the corresponding modes, obtaining classical results $p_i$ and $x_j$, from which she computes $g_i= p_i-\sum_{j\in N^{(i)}}\Omega_{\{i,j\}}  x_j$. \changes{At this step, the measurement of the quadrature operators is assumed to be noisy, which leads to a noisy estimation of $\hat g_i$. Details on the noise model and the link between the noise of the different operators are provided respectively in sections \ref{section: operators formalism} and \ref{sec: locally measuring the nullifiers}.}

    \item The next steps deviate from the protocol originally formulated in \cite{takeuchi_resource-efficient_2019}. To allow for imperfect states and measurements, we must loosen the acceptance criteria. To solve this issue we impose that Alice probabilistically accepts a measurement result $g$ if it is sufficiently small, considering the inherent noise in her measurements. She accepts it with probability $\exp(-g^2/\epsilon^2)$, for constant $\epsilon>0$. She then computes $N_\text{pass}$, the total number of accepted measurements.
    
    \item If $N_\text{pass}\geq \left(1-f\right)nN_\text{test}$, the protocol succeeds, and Alice randomly keeps $k$ registers that have not been measured yet. \changes{These registers form the state $\hat \rho_\text{tgt}$ which she can use for further application.} Here, $f$ is the failure threshold, defining the maximum allowable percentage of measurement rejections while still accepting the state at the end of the protocol. The protocol then guarantees that the corresponding state $\hat \rho'$ is close to the ideal graph state. Otherwise, the protocol fails and nothing can be inferred from the remaining states.
\end{enumerate}

To ensure a satisfactory certification protocol, it must adhere to two properties (see, {\it e.g.} \cite{markham_simple_2018}), that will be defined more rigorously later:
\begin{itemize}
    \item {\bf Completeness:}  When Bob behaves honestly, Alice should accept the state with a high probability, that is, states close to the ideal graph state should pass the test with high probability.
    \item {\bf Soundness:} The probability to accept a state that should not pass the protocol, due to being too distant from the ideal graph state, should be low.
\end{itemize}

\subsection{Modelling the POVMs}\label{section: operators formalism}
A key part for understanding what is happening during the protocol, in particular to show security, is to provide a complete expression for the Positive Operator-Valued Measure (POVM) associated with different parts of the protocol. Given the consideration of noisy measurements, a careful modeling of the process is necessary. In the protocol, whenever a state is selected for measurement or retention, it is chosen uniformly at random. For convenience, we instead assume that measurements are performed on the first registers and that the initial state $\hat\rho$ is permutation invariant, which is equivalent. This choice is reflected in Figure~\ref{fig: protocol}. For a fixed register, we can diagonalize the operator $\hat g_i$
\begin{equation}
    \hat g_i=\int g \hat E^{(i)}_g dg,
\end{equation}
where $\hat E_g^{(i)}$ is the projector onto the eigenspace of $\hat g_i$ associated to the eigenvalue $g$. To model noisy measurements of $\hat g_i$, we introduce a Gaussian with width $\delta_i$. The POVM associated with obtaining outcome $g$ is then defined as
\begin{equation}
    \hat E_{g,\delta_i}^{(i)}=\frac{1}{\delta_i\sqrt{\pi}}\int dg' \exp(-\frac{(g-g')^2}{\delta_i^2})\hat E_{g'}^{(i)}.
\end{equation}
Subsequently, in the post-processing step, an outcome $g$ is accepted with probability $\exp(-g^2/\epsilon^2)$. The POVM associated with accepting the measurement is given by
\begin{align}
    \hat P^{(i)}&=\int dg \exp(-\frac{g^2}{\epsilon^2})\hat E_{g,\delta_i}^{(i)}\notag\\
    &=\frac{\epsilon}{\sqrt{\epsilon^2+\delta_i^2}} \int dg' \exp(-\frac{g'^2}{\epsilon^2+\delta_i^2}) \hat E_{g'}^{(i)}.
\end{align}
Now, considering multiple registers, an index $j$ is introduced to denote operators acting on the $j$-th register. Consequently, we denote by $\hat P^{(i)}_j$ the corresponding POVM. Under the assumption that the state is permutation invariant, allowing the measurement to be assumed to be conducted on the first registers, the POVM corresponding to the acceptance of the entire state by the protocol is expressed as follows
\begin{widetext}
    \begin{equation}
        \hat\Pi_\text{acc}=\sum_{\substack{a_{i,j}=0,1\\1\leq i\leq n~\&~ 1\leq j\leq N_\text{test}\\ \sum a_{i,j}\geq (1-f)nN_\text{test}}} \prod_{i=1}^n\prod_{j=1}^{N_\text{test}}\left(\hat P^{(i)}_{(j-1)n+i}\right)^{a_{i,j}}\left(\1-\hat P^{(i)}_{(j-1)n+i}\right)^{1-a_{i,j}}.
    \end{equation}
\end{widetext}

To understand this expression, one can focus on the terms following the product symbols. The exponents, being either $0$ or $1$, based on the value of $a_{i,j}$, selects one of the two factors: $\hat P^{(i)}_{(j-1)n+i}$ or $\1-\hat P^{(i)}_{(j-1)n+i}$. These represent either the success ($a_{i,j}=1$) or failure ($a_{i,j}=0$) of the $\hat g_i$ test on register $(j-1)n+i$. The summation is then performed over all possible values of $a_{i,j}$, retaining only those corresponding to a sufficiently high number of successes ($\sum a_{i,j}\geq (1-f)nN_\text{test}$). Following these measurements, one gets a remaining state $\hat \rho_r$ on the last $N_\text{total}-nN_\text{test}$ registers. Assuming a successful test, Born's rules dictate that
\begin{equation}
    \hat \rho_r=\frac{\Tr_{nN_\text{test}}(\hat \rho \hat\Pi_\text{acc})}{\Tr(\hat\rho\hat\Pi_\text{acc})},
\end{equation}
where the partial trace $\Tr_{nN_\text{test}}$ is taken over the first $nN_\text{test}$ measured registers. The objective of the certification is to ensure that this state primarily consists of approximate copies of the ideal graph state. This is equivalent to satisfying the nullifier conditions. The POVM encoding the probability for a state to pass all nullifier states on register $i$ is given by
\begin{equation}
    \hat \Pi_i=\hat P^{(1)}_i\times \cdots\times \hat P^{(n)}_i.
\end{equation}
We aim for the state $\hat\rho_r$ to pass all nullifier tests for the largest number of modes. This leads to the consideration of the POVM
\begin{equation}
    \hat\Pi_\text{good}^N=\sum_{\substack{i_1,\dots,i_r\in \{0,1\}\\ i_1+\dots+i_r\geq N}} \prod_{k=1}^{N_r} \hat \Pi_k^{i_k}(\1-\hat \Pi_k)^{(1-i_k)},
\end{equation}
determining whether there are at least $N$ registers that pass all nullifier tests. For a fixed $N$, the probability that at least $N$ registers satisfy all nullifier conditions is given by
\begin{equation}
    \Tr(\hat\Pi_\text{good}^N\hat \rho_r)=\frac{\Tr((\hat\Pi_\text{acc}\otimes\hat\Pi_\text{good}^N)\hat\rho)}{\Tr(\hat \rho\hat\Pi_\text{acc})}.
\end{equation}
Here, $\Tr((\hat\Pi_\text{acc}\otimes\hat\Pi_\text{good}^N)\hat\rho)$ is the joint probability that the initial state is both accepted by the protocol and that $N$ systems remaining would satisfy all the nullifier conditions, while $\Tr(\hat \rho\hat\Pi_\text{acc})$ is the {\it a priori} probability that the initial state is accepted by the protocol.

\subsection{Locally measuring the nullifiers}\label{sec: locally measuring the nullifiers}
The POVM $P^{(i)}$ has been introduced to represent the imperfect measurement of the nullifier $\hat g_i$. However, in practical scenarios, $\hat g_i$ is not directly measured. Instead, local measurements are performed by evaluating $\hat p_i$ and $\hat x_j$ on the corresponding modes, followed by computing the linear combination $g_i= p_i-\sum\limits_{j\in N^{(i)}}\Omega_{\{i,j\}} x_j$. In this subsection, our goal is to establish that these two different perspectives yield the same mathematical representation. In particular, we verify that noisy versions of the local measurements correspond to the noisy nullifier measurement modeled above. To confirm this, we will consider the simplest non-trivial case and show that the POVMs associated with both viewpoints are equal. Let us consider a two-mode graph state with a nullifier given by
\begin{equation}
    \hat g=\hat p_1-\hat x_2.
\end{equation}
The operator $\hat g$ can be diagonalized as $\hat g= \int \hat P_g dg$ , with
\begin{equation}
    \hat P_g =\int d p  ~d x ~\delta \big[g-(p-x)\big]\ketbra{p}\otimes \ketbra{x},
\end{equation}
$\hat P_g$ being the projector on the eigenspace of $\hat g$ with eigenvalue $g$. If we consider the measurement of $\hat g$ as noisy, with a Gaussian noise of width $\delta$, the POVM associated with the measurement outcome $g$ is given by
\begin{equation}
    \hat E_g=\frac{1}{\delta\sqrt{\pi}}\int dg'~\exp[-\frac{(g-g')^2}{\delta^2}]\hat P_{g'},
\end{equation}
where the normalization constant is determined by the relation $\int dg ~\hat E_g=\1$. On the other hand, if we consider the measurement of $\hat p_1$ and $\hat x_2$ as noisy and combine their outcomes to `simulate' the measurement of $\hat g$, the corresponding POVM associated with the outcome $g$ is given by
\begin{equation}
    \hat M_g=\int dp~dx~\delta\big[g-(p-x)\big] \hat E^{\hat p}_p \otimes \hat E^{\hat x}_x,
\end{equation}
where $\hat E^{\hat x}_x =\frac{1}{\mu\sqrt{\pi}}\int dx' \exp[-\frac{(x-x')^2}{\mu^2}]\ketbra{x'}$ and $\hat E^{\hat p}_p =\frac{1}{\nu\sqrt{\pi}}\int dp' \exp[-\frac{(p-p')^2}{\nu^2}]\ketbra{p'}$ are the POVMs associated with the noisy measurement of $\hat x$ and $\hat p$, respectively. Assuming that $\delta^2=\nu^2+\mu^2$, a straightforward computation reveals that
\begin{equation}
    \hat M_g=\hat E_g.
\end{equation}

Born's rule dictates that for destructive measurement (the measured registers are not kept), the quantum state over the remaining part is solely determined by the measurement POVM. Therefore, both pictures indeed represent the same process.

\section{Security of the certification protocol}\label{sec: security of the certification protocol}
With all the definitions in place, we present the core statements of our contribution. We have opted to provide the most general statement first, retaining the largest number of free parameters. In practice, one can then fix relations between these parameters so that the different bounds are more manageable, depending on the task at hand. Simplified expression will be presented in section \ref{sec: special cases}. To help the reader, we repeat the definitions of all the parameters that characterize the certification protocol: 
    \begin{itemize}
        \item $n$: the number of modes in the target CV graph state.
        \item $N_\text{total}$: the total number of registers (requested copies).
        \item $N_\text{test}$: the total number of times each nullifier $\hat g_i$ is measured on a different register. Since there are $n$ different nullifiers, this results in a total number of $n N_\text{test}$ measurements.
        \item $\mu:= N_\text{total}/N_\text{test}$. Since we need to have more total registers than the numbers of measured ones, we need $\mu>n$.
        \item $f$: the threshold on the fraction rejection after each register measurement. To accept the state, the percentage of failure should be less than $f$.
        \item $\delta$: the characteristic noise of the nullifier measurements.
        \item $\epsilon$: the characteristic width of the probability to accept a measurement outcome $g$.
        \item $\nu$: a free parameter that appears in Serfling's bound.
    \end{itemize}

\subsection{Completeness and Soundness}\label{sec: theorems}
    
We first address the easiest question: completeness. 
As described earlier, completeness corresponds to showing that states close to the ideal graph state will be accepted by the protocol under reasonable choices of the parameters. We provide two formal versions of this statement. The first is for a specific approximation of the target ideal graph states defined in (\ref{eq: graph state finite squeezing})- the standard Gaussian finite squeezed version. The second definition of completeness works for any state with close enough overlap (see (\ref{eq: overlap})). 

First, if we assume that the incoming state is made of identical copies of state (\ref{eq: graph state finite squeezing}).     \begin{Thm}\label{thm: completeness gaussian}{\bf (Completeness for finitely squeezed approximate graph states)}

     The probability that the state $\ket{G^\sigma}$ passes the nullifier test $\hat g_i$ is:
     \begin{align}
        P_{\text{null}}&=\bra{G^\sigma} \hat P^{(i)}\ket{G^\sigma}=\left(1+\left(\delta^2+\frac{1}{\sigma^2}\right)\frac{1}{\epsilon^2}\right)^{-1/2}.
     \end{align}
     The state is accepted if:
     \begin{equation}
         P_{\text{acc}}=\sum_{k\geq (1-f)nN_{\text{test}}} \binom{nN_{\text{test}}}{k}P_{\text{null}}^k(1-P_{\text{null}})^{nN_{\text{test}}-k}.
     \end{equation}
So with $\epsilon^2\gg (\delta^2+1/\sigma^2)$, the probability $P_\text{acc}$ can be as close as desired to $1$, for any level of noise $\delta$ and required squeezing $\sigma$.    
\end{Thm}

\begin{preuve}
    This theorem is obtained by straightforward integrals computations.
\end{preuve}
Another version of the completeness is based on the overlap $O^\Delta$. We show that given that it is close to $1$, then the state is accepted with high probability. 
\begin{Thm}\label{thm: completeness gen fid}{\bf (Completeness)}
    For a state $\hat \rho$ satisfying $O^\Delta>1-\eta$, the state $\hat\rho^{\otimes N_\text{total}}$ will be accepted by the protocol with high probability given that $\epsilon \gg (\delta+\Delta)$ and $\eta\ll 1$. More precisely, the probability that a nullifier measurement is accepted can be lower-bounded as
    \begin{equation}
        P_\text{null}\geq e^{-(\delta+\Delta)/\epsilon}\left(1-\eta-2e^{-\epsilon/(\mu+\Delta)}\right).
    \end{equation}
    As the probability to accept the state is given as in the previous theorem by 
     \begin{equation}
         P_{\text{acc}}=\sum_{k\geq (1-f)nN_{\text{test}}} \binom{nN_{\text{test}}}{k}P_{\text{null}}^k(1-P_{\text{null}})^{nN_{\text{test}}-k}.
     \end{equation}
     By having $P_\text{null}$ sufficiently close to one, one can have $P_\text{acc}$ to be as close as one wants to $1$.
\end{Thm}

\begin{preuve}
    See \ref{proof: completeness gen fid} for a detailed derivation. The intuition of the proof is the following. We prove in proposition \ref{pro: concentration on g} that under the assumption $O^\Delta\geq 1-\eta$, the measurements of each nullifier are concentrated around zero. Hence, with a high probability, they will be accepted.
\end{preuve}
One can note, that the overlap $O^\Delta$ of the Gaussian state $\ket{G^\sigma}$ can explicitly be computed, thus one could also apply thm.~\ref{thm: completeness gen fid} to this type of state. However, the bound that would be obtained would be less sharp than the one given in thm.~\ref{thm: completeness gaussian}. It is useful to compare the results obtained here with the protocol outlined in \cite{takeuchi_resource-efficient_2019}. In that protocol, completeness was considered trivial since when the verifiers received copies of the ideal state, the protocol would accept it with unit probability. However, in our scenario, due to the presence of noise and probabilistic acceptance, the parameters need to be carefully tuned to ensure that the acceptance probability is sufficiently high. 

Before being able to give a statement about the soundness of the protocol, we introduce an important intermediary statement that contains all the technical machinery. It is mainly based on Serfling's bound \cite{serfling_probability_1974}, which is the main technical ingredient of security proof in \cite{takeuchi_resource-efficient_2019}. We should also give a quick summary of the differences in the proof compared to \cite{takeuchi_resource-efficient_2019}. Our proofs are essentially based on the same ideas. However, we adapted them to take the noisy measurements, and noisy input state into account which is essential to have a meaningful protocol (without this the protocol would never accept, for example). This requires careful tracing of all the steps involved, and adapting them to the noisy case, which is non-trivial. We present several forms of statements related to soundness, which offers more flexibility in how we choose the free parameters and how they relate to each other, as well as more visibility in comparing different figures of merit for soundness.
 
 \begin{Thm}\label{thm: serfling}
    {\bf (Serfling's bound)} For any initial state $\hat\rho$ over all requested $N_\text{total}$ registers, for $N=nN_\text{test}(\mu/n -1-\mu f-\mu\nu)$ we have,
    \begin{equation}\label{eq: serfling product}
        \Tr((\hat\Pi_\text{acc}\otimes[\1-\hat\Pi_\text{good}^N])\hat\rho)\leq n \exp(-\frac{\nu^2N_\text{test}}{1+\frac{1}{\mu-n}}),
    \end{equation}
    subject to the constraint $\mu/n\geq 1+\mu f+\mu \nu$ to ensure $N$ is positive. We also can impose $\nu^2 N_\text{test}>2\ln(n)$ to ensure the bound is less than $1$, providing useful constraints. The bound can be rephrased as a conditional statement
    \begin{equation}\label{eq: serfling conditionnal}
        \Tr(\hat\Pi_\text{good}^N\hat \rho_r)\geq 1-\frac{n \exp(-\frac{\nu^2N_\text{test}}{1+\frac{1}{\mu-n}})}{\Tr(\hat\Pi_\text{acc}\hat \rho)},
    \end{equation}
    where $\hat \rho_r$ is the state over the remaining registers after a successful protocol.
\end{Thm}

\begin{preuve}
    See \ref{proof: operator bound} for a detailed derivation. The proof builds on the approach of \cite{takeuchi_resource-efficient_2019} using Serfling's bound, providing a more extensive and detailed derivation that accounts for the operator formulation, in particular explicit application of the POVM for the noisy measurement as described in section \ref{section: operators formalism}.
    The intuitive idea is that Serfling's \cite{serfling_probability_1974} concentration inequality allows us to conclude that if a large number of measurements have been accepted, then a large fraction of additional measurements would also likely be accepted. 
\end{preuve}

The first inequality (\ref{eq: serfling product}) of theorem \ref{thm: serfling} can be understood as follows. We recall that the POVM $\hat\Pi_\text{acc}$ flags whether a state is accepted or not by the protocol, while $\hat\Pi_\text{good}^N$ tells from the remaining unmeasured states, whether at least $N$ of them would successfully pass all nullifier tests. This inequality establishes a limit on the probability of a state being accepted when the majority of the remaining registers are predominantly `bad' states. The second inequality (\ref{eq: serfling conditionnal}) is a conditional statement that assumes a successful measurement step and the decision to accept the remaining state. The formula then gives a lower bound on the probability that the remaining registers indeed contain at least $N$ `good' states. While this statement holds theoretical significance, its practical utility may be limited. Indeed, it gives information on the number of `good' states among the large collection of unmeasured ones. However, it leaves unaddressed questions about the quality of a subset of states selected for further applications, say, retaining $k$ states. What can be said about the quality of these states? This is answered by the following statement.
\begin{widetext}

    \begin{Thm}\label{thm: soundness}
    {\bf (Soundness)} For any initial state $\hat \rho$ and integer $k\leq N_r:= (\mu-n)N_\text{test}$ we have:
     \begin{align}\label{eq: soundness non-conditionnal}
        &\Tr(\hat \Pi_\text{acc}\otimes[\1-\hat\Pi_1\times\cdots\times\hat \Pi_k]\hat \rho)\leq  1-\left(1-\frac{knN_\text{test}\mu(f+\nu)}{(\mu-n)N_\text{test}-k+1}\right)\times\left [1-n\exp(-\frac{\nu^2 N_\text{test}}{1+\frac{1}{\mu-n}})\right],
    \end{align}
    where, as before we are under the constraint that $\mu/n\geq 1+\mu f+\mu\nu$. A conditional statement can be formulated as follows
        \begin{align}\label{eq: soundness conditionnal}
        &\Tr(\hat\Pi_1\times\cdots\times\hat\Pi_k \hat \rho_r)\geq \left(1-\frac{knN_\text{test}\mu(f+\nu)}{(\mu-n)N_\text{test}-k+1}\right)\times\left(1-\frac{n\exp(-\frac{\nu^2 N_\text{test}}{1+\frac{1}{\mu-n}})}{\Tr(\hat\Pi_ \text{acc}\hat\rho)}\right),
    \end{align}
    where $\hat \rho_r$ is the state over the remaining register after a successful protocol.
\end{Thm}
\end{widetext}

\begin{preuve}
    Refer to \ref{proof: soundness} for a detailed derivation. This result follows quite straightforwardly from the previous theorem \ref{thm: serfling}, primarily relying on an intuitive probabilistic argument. The operators $\hat\Pi_\text{good}^N$ ascertain whether there are at least $N$ good states among the remaining ones. If now, one wants to know if the first $k$ are good, one can obtain a lower bound by considering the following: the probability that $k$ states are good if chosen uniformly at random among $N_r$ state, where $N$ of them are good is given by the quotient of the binomial coefficient $\binom{N}{k}/\binom{N_r}{k}$. Simplifying this quotient and combining it with the probability obtained in the previous theorem yields the results.
\end{preuve}
The first inequality (\ref{eq: soundness non-conditionnal}) in the theorem is structured similarly to the one in the previous theorem. It bounds from above the probability that a state is accepted by the protocol while still failing to pass the nullifier test on the $k$ registers that have been kept. Similarly, as before, a conditional statement can also be formulated (\ref{eq: soundness conditionnal}), bounding from below the probability that the nullifier test on the $k$ kept registers will succeed if we assume that the state just got accepted by the protocol. 

Note that the above form is rather general and does not use a specific choice of definition of closeness. However, for different applications (see section \ref{sec: application}), it is useful to do so. To this end, we define a soundness statement in terms of overlap (\ref{eq: overlap}).
\begin{widetext}

    \begin{Thm}\label{thm: fidelity bound general k systems}
        {\bf (Overlap for $k$ systems)} If the protocol is executed with an initial state $\hat \rho$ and in the case that it is successful ({\it i.e.} the state is accepted by the POVM $\hat\Pi_\text{acc}$) and randomly a $k$ register state $\rho'$ is retained, the overlap can be bounded as follows
        \begin{equation}
            O_k^{\sqrt{\epsilon^2+\delta^2}}\geq \left(1-\frac{knN_\text{test}\mu(f+\nu)}{(\mu-n)N_\text{test}-k+1}\right)\left(1-\frac{n\exp(-\frac{\nu^2 N_\text{test}}{1+\frac{1}{\mu-n}})}{\Tr(\hat\Pi_\text{acc}\hat\rho)}\right).
        \end{equation}
        Where $\epsilon$ is the parameter governing the probability of accepting a measurement result $e^{-g^2/\epsilon^2}$. And $\delta$ is the classical noise when performing a $\hat g_i$ measurement. If $\hat g_i=\hat p_i-\sum\limits_{j\in N^{(i)}}\Omega_j \hat x_j$ is measured with $\hat p_i$ and $\hat x_j$ operators having respective noise $\nu$ and $\mu$, then $\delta =\max\limits_{i} \delta_i$, with $\delta_i^2= \nu^2+\mu^2\sum\limits_{j\in N^{(i)}} \Omega_j^2 $.
    \end{Thm}    
\end{widetext}

\begin{preuve}
    See \ref{proof: fidelity bound k systems} for detailed proof. The underlying ideas are the same as the one of the second proof presented in the preceding theorem. Essentially, it revolves around recognizing the striking similarity between the expanded expressions of $\Tr(\hat\Pi_1\times\cdots\times\hat\Pi_k \hat \rho_r)$ and $O_k^{\sqrt{\epsilon^2+\delta^2}}$. It is then straightforward to verify that $O_k^{\sqrt{\epsilon^2+\delta^2}}$ is indeed greater than or equal to $\Tr(\hat\Pi_1\times\cdots\times\hat\Pi_k \hat \rho_r)$.
\end{preuve}
One has to keep in mind that theorem \ref{thm: fidelity bound general k systems} is also a conditional statement where it is assumed that the state got accepted by the protocol. \changes{This theorem gives information of the form $O^\Delta\geq 1-\eta$ on the quantum state and explicitly expresses the parameters $\Delta$ and $\eta$. In section \ref{sec: application} different applications are discussed and use the assumption $O^\Delta\geq 1-\eta$. If our protocol is used to verify the state, then this theorem provides an explicit expression for $\Delta=\sqrt{\delta^2+\epsilon^2}$ and $\eta$.} As the case where $k=1$ will be extensively used in section \ref{sec: application}, we formulate the immediate corollary, which corresponds to taking $k=1$ in the previous theorem.

\begin{Cor}\label{thm: fidelity bound general}
    {\bf (Overlap for $1$ system)} If the protocol is executed with an initial state $\hat \rho$, and is successful ({\it i.e.} the state is accepted by the POVM $\hat\Pi_\text{acc}$), for a single register output the overlap can be bounded as follows
    \begin{equation}
        O^{\sqrt{\epsilon^2+\delta^2}}\geq \left(1-\frac{n\mu(f+\nu)}{\mu-n}\right)\left(1-\frac{n\exp(-\frac{\nu^2 N_\text{test}}{1+\frac{1}{\mu-n}})}{\Tr(\hat\Pi_\text{acc}\hat\rho)}\right).
    \end{equation}
    We recall that $\epsilon$ is the parameter governing the probability of accepting a measurement result and $\delta$ is the size of the classical noise when performing a $\hat g$ measurement. Specifically, if one measure $\hat g_i=\hat p_i-\sum\limits_{j\in N^{(i)}}\Omega_j \hat x_j$ by measuring the corresponding $\hat p_i$ and $\hat x_j$ operators with respective noise $\nu$ and $\mu$, then $\delta =\max\limits_{i} \delta_i$, with $\delta_i^2= \nu^2+\mu^2\sum\limits_{j\in N^{(i)}} \Omega_j^2 $.
\end{Cor}

\begin{preuve}
    The proof is obtained by setting $k=1$ in thm.~\ref{thm: fidelity bound general k systems}.
\end{preuve}

\subsection{Special cases}\label{sec: special cases}
In the result we have given in section \ref{sec: theorems}, we opted for the most general scenario, introducing numerous free parameters that allow for flexibility but also complicate the interpretation. Consequently, it becomes challenging to discern optimal choices and understand how the bounds scale when varying the number of used copies. This section addresses these challenges by presenting and discussing different possibilities. 

By constraining the parameters, for example, one can arrive at a formulation similar to  \cite{takeuchi_resource-efficient_2019}, by setting
\begin{align}
    \mu =2n && N_\text{test}=\left\lceil 5n^4\frac{\log n}{32}\right\rceil && \nu =\frac{\sqrt{c}}{n^2} && f=\frac{1}{2n^2},
\end{align}
where $c$ is some constant. With these choices, the soundness bound (eq.~\ref{eq: soundness non-conditionnal}) for $k=1$ becomes
\begin{align}\label{eq: soundness original}
    \Tr(\hat\Pi_\text{acc}(\1-\hat\Pi_1)\hat\rho)&\leq1- \left(1-\frac{2\sqrt{c}+1}{n}\right)\left(1-n^{1-5c/64}\right)\notag\\
    &\leq \frac{2\sqrt{c}+1}{n}+n^{1-5c/64}.
\end{align}
The constraints formulated in the theorem read in this case
\begin{align}
    c<\frac{(n-1)^2}{4} && c>\frac{64}{5}
\end{align}
Combining these constraints reveals that they imply $n\geq 9$. First notice that our statements are formulated slightly differently from \cite{takeuchi_resource-efficient_2019} even still. However, in equation (\ref{eq: soundness original}) we recognize $1-2(\sqrt{c}+1)/n$ which was the fidelity of the state and $n^{1-5c/64}$ which was the error probability in the formulation of \cite{takeuchi_resource-efficient_2019}. 
The main thing we notice with this choice is that the only remaining free parameter is $n$ (with $c$ also free but constrained by $n$, and one will consider the value of $c$ that optimizes the bound). Therefore, the only way to tighten the bound in (\ref{eq: soundness original}) is by increasing the value of $n$. This poses a problem when applying this certification protocol to check specific graph states of fixed size, and the requirement $n\geq 9$ prevents the use of this protocol for the certification of small graph states. This is what motivated the general formulation. Assuming a fixed size $n$ for the graph state, we introduce a new variable $\lambda$ to control the tightness of the bound. With the following parameter choices
\begin{align}\label{eq: choice with lambda}
    \mu=2n && N_\text{test}=\left\lceil (2\ln n+2\ln \lambda)\lambda^2\right\rceil && \nu=1/\lambda && f=1/\lambda,
\end{align}
the bound reduces to
\begin{align}\label{eq: soundness lambda}
    \Tr(\hat\Pi_\text{acc}(\1-\hat\Pi_1)\hat\rho)&\leq1- \left(1-\frac{4n}{\lambda}\right)\left(1-\frac{1}{\lambda}\right)\notag\\
    &\leq\frac{4n+1}{\lambda},
\end{align}
with the only non-trivial condition being $\lambda > 4n$. In this case, one can arbitrarily increase the parameter $\lambda$ to sharpen the bound.  If we consider keeping $k$ registers at the end of the protocol, the same choice of parameters yields

\begin{align}\label{eq: soundness lambda k}
    &\Tr(\hat \Pi_\text{acc}\otimes[\1-\hat\Pi_1\times\cdots\times\hat \Pi_k]\hat \rho)\notag\\
    &~~~~~~\leq 1- \left(1-\frac{4n}{\lambda}\frac{k}{1-\frac{k}{\lambda^2}}\right)\left(1-\frac{1}{\lambda}\right).
\end{align}

Similar computations can be made to adapt and simplify the other equations presented in the result section \ref{sec: theorems}.

\changes{It is important to notice that both our protocol and the one in \cite{takeuchi_resource-efficient_2019} work in almost the same way. As such our choice of parameter in terms of $\lambda$ can also be done for their protocol. The fact that our protocol takes the noise and the state imperfection into account does not change how it runs: the right-hand-side of equation (\ref{eq: soundness lambda}) and (\ref{eq: soundness lambda k}) do not depend on the noise parameter $\delta$, nor on the squeezing level $\sigma$. Instead the operators $\hat \Pi_i$ depends on $\delta$. All of these means that the number of resources necessary for soundness of the verification is independent of the measurement noise. However completeness is not possible in the noiseless case since no physical state will ever pass the test. A more indepth comparision between this work and others, including \cite{takeuchi_resource-efficient_2019} can be found in section \ref{Sec Comparison}. For now we are more interested in exploring how our bounds work in practice with explicit numbers.}

\subsection{Explicit numbers}
In this section, we aim to analyze the practical implications of the formulas (\ref{eq: soundness lambda}) and (\ref{eq: soundness lambda k}) presented in section \ref{sec: special cases}. By making specific parameter choices, we can gain insight into the required number of copies, noise levels, squeezing, and other factors in realistic scenarios. Let us recall that if a state successfully passes the protocol and a single register is kept, the joint probability that the state passes the protocol, while the retained state is far from a graph state is bounded by
\begin{equation}\label{eq: soundness lambda v2}
    J=\Tr(\hat\Pi_\text{acc}(\1-\hat\Pi_1)\hat\rho)\leq\frac{4n+1}{\lambda},
\end{equation}
with the choice
\begin{align}
    N_\text{total}=2nN_\text{test} && N_\text{test}=\lceil (2\ln n+2\ln \lambda)\lambda^2\rceil && f=1/\lambda.
\end{align}
Assuming a joint probability $J$ less than $0.1$ , we can isolate $\lambda$ in eq.~(\ref{eq: soundness lambda v2}) and get

\begin{equation}
    \lambda=\frac{4n+1}{J}.
\end{equation}

Subsequently, we can determine the values of $N_\text{test}$ and $N_\text{total}$ by substituting $\lambda$ into their definitions. Remarkably, these expressions remain independent of measurement noise $\delta$, measurement threshold $\epsilon$, squeezing $\sigma$, and the failure fraction $f$. Moving forward, if we aim for a state that passes the protocol with a probability of $P_\text{acc}=0.9$, we can compute the probability that a single measurement should be successful ($P_\text{stab}$). \changes{Both quantities $P_\text{acc}$ and $P_\text{stab}$ are functions of the parameters defining the protocol (see section \ref{sec: security of the certification protocol}).} Utilizing the relation
\begin{equation}
    P_\text{acc}=\sum_{k=(1-f)nN_\text{test}}^{nN_\text{test}}\binom{nN_\text{test}}{k}P^k_\text{stab}(1-P_\text{stab})^{nN_\text{test}-k}.
\end{equation}
we can numerically evaluate $P_\text{stab}$ by using the previously obtained value for $N_\text{test}$  and the relation $f=1/\lambda$. Furthermore, we have the relation (thm. \ref{thm: completeness gaussian})
\begin{equation}
    P_\text{stab}=\frac{\epsilon\sigma}{\sqrt{1+(\epsilon^2+\delta^2)\sigma^2}},
\end{equation}
which translate to
\begin{equation}
    \epsilon^2=\frac{P_\text{stab}^2}{1-P_\text{stab}^2}\left(\frac{1}{\sigma^2}+\delta^2\right).
\end{equation}
This reveals that the value of $\epsilon$ is constrained by both the noise parameter $\delta$ and the squeezing level $\sigma$. In essence, demanding excessively high squeezing does not significantly reduce the value of $\epsilon$. A reasonable magnitude for $\sigma$ is approximately $1/\delta$, leading to
\begin{equation}
    \epsilon=\frac{\sqrt{2}P_\text{stab}}{\sqrt{1-P_\text{stab}^2}}\delta.
\end{equation}
Now, we can compute the numbers for different graph state sizes ($n$), as shown in Table \ref{tab: number one register}
\begin{table}[H]
    \centering
    \begingroup
        \setlength{\tabcolsep}{6pt} 
        \renewcommand{\arraystretch}{1.5} 
        \begin{tabular}{|c||c|c|c|c|c|}
        \hline
            $n$ & $\lambda$ & $N_\text{test}$ & $N_\text{tot}$ & $P_\text{stab}$ & $\epsilon/\delta$\\
            \hline\hline
            $1$ & $50$ & $20~000$ & $40~000$ & $0.981$ & $7$\\
            \hline
            $2$ & $90$ & $84~000$ & $340~000$ & $0.989$ & $10$\\
            \hline
            $5$ & $210$ & $620~000$ & $6~200~000$ & $0.995$ & $15$\\
            \hline
            $10$ & $410$ & $3\times10^6$ & $6\times10^7$ & $0.998$ & $20$\\
            \hline
            $100$ & $4~020$ & $4\times10^8$ & $8\times10^{10}$ & $0.9998$ & $63$\\
            \hline
        \end{tabular}
    \endgroup
    \caption{Table of the values of the parameters $\lambda$, $N_\text{test}$, $N_\text{tot}$, $P_\text{stab}$ and $\epsilon/\delta$ for various graph state sizes $n$ for the joint probability that a state is wrongly accepted to be less than $0.1$.}
    \label{tab: number one register}
\end{table}

\section{Application}\label{sec: application}
In this section, we operate under the assumption that our protocol has been successfully executed, resulting in the attainment of a state represented by $\hat\rho$. Corollary \ref{thm: fidelity bound general} shows that the state satisfies the condition
\begin{equation}
    O^\Delta\geq 1-\eta,
\end{equation}
where the expressions of $\eta$ and $\Delta$ are given by the corollary. In practical scenarios, both $\Delta$ and $\eta$ will be close to $0$. Acknowledging the theoretical nature of this bound, our objective is to delve into its implications for practical applications.

\subsection{Consequence on nullifier measurement}\label{sec: consequence on nullifier measurement}
In this section, we aim to analyze the practical consequences of assuming $O^\Delta\geq 1-\eta$. As we will see, it
provides a constraint on the distribution of the nullifier
measurement outcomes. To develop intuition, we first present a negative result.
\begin{Pro}\label{pro: variance not bounded}
    The variance of the measurement of $\hat g_i$ can be as high as one wants, and can even be infinite.
\end{Pro}

\begin{preuve}
See \ref{proof: variance not bounded} for a detailed proof. Intuitively, it works as follows. To get a state that satisfies $O^\Delta\geq 1-\eta$, one can simply take a state whose measurement results for the nullifiers are exactly $0$ with probability $1-\eta$. For the remaining probability $\eta$, the distribution can be anything, and thus as wide as one wants.
\end{preuve}

While we cannot establish a bound on the variance, we can still derive a concentration inequality.
\begin{Pro}\label{pro: concentration on g}
    For any $x>0$, we have 
    \begin{equation}
       \P[\abs{g_i}\leq x]\geq 1-\eta-e^{-x^2/\Delta^2},
    \end{equation}
    which expresses the fact that the probability that the measured value of $\hat g_i$ is close to zero is bounded from below.
\end{Pro}

\begin{preuve}
    See \ref{proof: concentration on g}.
\end{preuve}
\changes{This bound has a very concrete physical interpretation. Assuming that $\eta\simeq 0$ is small and applying it to $x\lesssim\Delta$ we obtained $\P[\abs{g_i}\lesssim\Delta]\simeq 1$. The parameter $\Delta$ then gives the order of magnitude in the size of the nullifier measurement outcomes.} This establishes a bound on perfect measurement, but since the state certification protocol accounts for measurement noise, it is sensible to consider it in this case as well. Let $\delta_j$ represent a classical noise term, assumed Gaussian with width $\mu$, and independent of the distribution of the measurement of $\hat g_i$. We now wish to prove a concentration on $\abs{\hat g_i+\delta_j}$. Given the independence between the two random variables, we express the density of the sum as a convolution:
\begin{align}
    d\P[\hat g_i+\delta_j=s]&=\int dg d\P[\hat g_i=g]d\P[\delta_j=s-g]\notag\\
    &=\frac{1}{\mu\sqrt{\pi}}\int dg p(g) e^{-(s-g)^2/\mu^2},
\end{align}
where $d\P[\cdot]$ represents the density associated to the event $[\cdot]$ and $p(g)$ is the density associated to the measurement of $\hat g_i$. Three different bounds, based on slightly different approaches to bounding the terms, can be shown.
\begin{Pro}\label{pro: concentration g delta}
    For any $x>0$, we have 
    \begin{equation}\label{eq: concentration g delta V1}
        \P[\abs{\hat g_i+\delta_j}\leq x]\geq (1-\eta)\sqrt{1-\left(\frac{\mu}{\Delta}\right)^2}-e^{-\frac{-x^2}{\Delta^2-\mu^2}},
    \end{equation}
    \begin{equation}\label{eq: concentration g delta V2}
        \P[\abs{\hat g_i+\delta_j}\leq x]\geq 1-\eta-e^{-x^2/4\Delta^2}-\frac{2\mu}{x}e^{-x^2/4\mu^2},
    \end{equation}
    \begin{equation}\label{eq: concentration g delta V3}
        \P[\abs{\hat g_i+\delta_j}\leq x]\geq 1-\eta-\left(1+\frac{\mu+\Delta}{x}\right)e^{-\frac{x^2}{(\mu+\Delta)^2}},
    \end{equation}
    indicating that we bound from below the probability that the measured value of $\hat g_i$ with additional noise $\delta_j$ is close to zero.
\end{Pro}
\begin{preuve}
    See \ref{proof: concentration g delta}.
\end{preuve}

Upon analyzing the scaling of the three bounds for large values of $x$, it becomes evident that the last one (\ref{eq: concentration g delta V3}) is the most optimal in the regime of large $x$, where the probabilities are close to $1$. In practical scenarios where manipulating probabilities close to $1$ is crucial, this bound emerges as the most informative.

\subsection{Continuous variable quantum teleportation}
For the case $n=2$ and the non-trivial graph $G$ on two vertices (where the edge carries a weight of $1$), the corresponding graph state proves useful in continuous variable state teleportation. A notable characteristic of graph states employed in this protocol is the significant correlation observed in the measurement of different quadrature operators. Since $\hat g_1=\hat p_1-\hat x_2$ always evaluates to zero on the perfect graph state, we know that the measurement of $\hat p_1$ and $\hat x_2$ are perfectly correlated and should always yield the same result (the same goes for $\hat g_2=\hat p_2-\hat x_1$). In the case of non-ideal graph states, while this correlation is not perfect, it remains robust, as demonstrated in the preceding section. This allows us to perform a fairly good teleportation. Now consider a teleportation scheme involving an input mode described by the quadrature operators $\hat x_\text{in}$, $\hat p_\text{in}$. Teleportation is achieved via a two-mode continuous variable graph state described by the operators $\hat x_1$, $\hat p_1$, $\hat x_2$, and $\hat p_2$.  For simplicity, we assume a weight of one for the edge, leading to nullifiers 
\begin{align}
    \hat g_1=\hat p_1-\hat x_2 && \hat g_2= \hat p_2-\hat x_1.
\end{align}
One then performs a joint measurement on the input and first modes given by the operators
\begin{align}
    \hat g_a=\hat p_\text{in} -\hat x_1 && \hat g_b=\hat p_1-\hat x_\text{in}.
\end{align}
By doing so, one obtains the classical outcomes, $g_a$ and $g_b$. We then have the following relation between the operators
\begin{align}
    \hat x_2=\hat p_1-\hat g_1=\hat x_\text{in}+\hat g_b-\hat g_1=\hat x_\text{in}+g_b-\hat g_1\notag\\
    \hat p_2=\hat x_1+\hat g_2=\hat p_\text{in}-\hat g_a+\hat g_2=\hat p_\text{in}-g_a+\hat g_2.
\end{align}

Upon classically communicating the joint measurement outcome to mode $2$, a conditional displacement allows to obtain
\begin{align}
    \hat x_2\longrightarrow \hat x_2'=\hat x_2-g_b=\hat x_\text{in}-\hat g_1\notag\\
    \hat p_2\longrightarrow \hat p_2'=\hat p_2+g_a=\hat p_\text{in}+\hat g_2.
\end{align}

The teleported operators ($\hat x_2'$ and $\hat p_2'$) deviate from the original ones ($\hat x_\text{in}$ and $\hat p_\text{in}$) only by the nullifiers of the CV graph state. Under the assumption that $O^\Delta\geq 1-\eta$, we can assert with high probability that this deviation is small (as stated in Proposition \ref{pro: concentration on g}). In the presence of noisy measurements, where $g_a$ and $g_b$ are subject to Gaussian noise, the displaced operators are expressed with new noise terms
\begin{align}
    \hat x_2'=\hat x_\text{in}-\hat g_1+\delta_b\notag\\
    \hat p_2'=\hat p_\text{in}+\hat g_2+\delta_a.
\end{align}

Here, $\delta_a$ and $\delta_b$ are classical Gaussian noise terms centered at zero, with variance $\mu^2$. Proposition \ref{pro: concentration g delta} allows us to conclude that, once again, the deviation between the initial and teleported operators is small, signifying a good approximate teleportation.

\subsection{Measurement-based quantum computing (MBQC)}

To perform quantum computation, Measurement-Based Quantum Computing (MBQC) diverges from the conventional gate-based model \cite{menicucci_universal_2006}. Instead of successively applying gates to a quantum state, MBQC involves a series of measurements in well-chosen bases on a large entangled state. Graph states can be employed in this context, and we thus want to understand how imperfections in the state, arising from the protocol application, impact MBQC performance. \\

In the gate-based model, achieving universal single-mode computation requires the ability to apply gates such as $\hat X(s)=e^{is\hat x}$, $e^{is\hat x^2}$ and $e^{is\hat x^3}$ along with the Fourier transform $\hat F=e^{i\pi(\hat x^2+\hat p^2)/4}$. Extending this to universal multi-mode computation involves adding the entangling gate $\hat {CZ}_{i,j}(s)=e^{is\hat x_i\hat x_k}$. In the subsequent discussion, we revisit the principles of MBQC within the context of continuous variable systems, drawing inspiration from the presentation outlined in \cite{gu_quantum_2009}. We commence with a straightforward example inspired by quantum teleportation

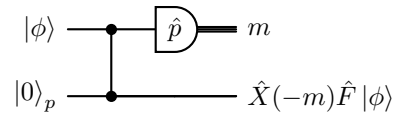
\begin{center}
    \begin{quantikz}
        \lstick{$\ket{\phi}$} & \ctrl{1} & \meterD{\hat p} & \cw  \rstick{$m$}\\
        \lstick{$\ket{0}_p$} & \control{} & \qw & \qw 
        \rstick{$\hat X(-m)\hat F\ket{\phi}$}\\
    \end{quantikz}   
\end{center}
The link between the two wires corresponds to the application of the entangling gate $\hat{CZ}(1)$. To see why this circuit indeed accomplishes what is written, let us consider its action on the quadrature operators $\hat x_1$, $\hat p_1$, $\hat x_2$, and $\hat p_2$. Recall that when going from Shrödinger's picture to the Heisenberg one, a state evolution $\ket{\psi'}=\hat U\ket{\psi}$ corresponds to the observable evolution $\hat{\mathcal O}'=\hat U^\dagger \hat{ \mathcal{ O}}\hat U$. This ensures that the expectation values are consistent in both pictures. 
\begin{align}
    \bra{\psi'}\hat{\mathcal O}\ket{\psi'}=\bra{\psi}\hat U^\dagger \hat{\mathcal O}\hat U\ket{\psi}=\bra{\psi}\hat{\mathcal O}'\ket{\psi}.
\end{align}
After the entangling gate, the new quadrature operators are expressed as follows
\begin{align}
    &\hat p_1'=\hat{CZ}(-1) \hat p_1 \hat{CZ}(1)=\hat p_1 +\hat x_2 &&\hat p_2'=\hat p_2 +\hat x_1\notag\\
    &\hat x_1'= \hat x_1 && \hat x_2'=\hat x_2.
\end{align}
Subsequently, the measurement step imposes $\hat p_1'=m$. Coupled with the initial state $\ket{0}_p$ verifying $\hat p_2=2$, we obtain
\begin{subequations}
    \begin{align}
        \hat x_2'&=\hat x_2=\hat p_1'-\hat p_1=m-\hat p_1=m+\hat F^\dagger \hat x_1 \hat F\notag\\
        &=\hat F^\dagger(\hat x_1+m)\hat F=\hat F^\dagger \hat X(m)\hat x_1\hat X(-m)\hat F\notag\\
        &=(\hat X(-m)\hat F)^\dagger \hat x_1 (\hat X(-m)\hat F)\\
        \hat p_2'&= \hat p_2+\hat x_1=\hat x_1=-\hat F^\dagger \hat p_1\hat F=-\hat F^\dagger \hat X(m)\hat p_1\hat X(-m)\hat F\notag\\
        &=(\hat X(-m)\hat F)^\dagger \hat p_1 (\hat X(-m)\hat F).
    \end{align}
\end{subequations}
This indeed represents the Heisenberg version of the Schrödinger evolution $\hat X(-m)\hat F\ket{\phi}$. This simple example, corresponds to the application of the identity gate, up to a known rotation ($\hat F$) and a translation ($\hat X(-m)$). This discrepancy between the desired output state and the one the protocol gives is unavoidable in the MBQC picture. One can then either, correct this difference by the suitable application of local rotation and translation, or one can modify the subsequently applied transformation to push the correction at the end. In the case of imperfect measurement and imperfect graph state, two noise terms are added: $\hat p_2\simeq 0$ (imperfect ancillary state; the dispersion is controlled by $O^\Delta\geq 1-\eta$) and $\hat p_1'= m+\delta$ (imperfect measurement; with Gaussian noise $\delta$).
\begin{align}
    \hat x_2'=(\hat X(-m)\hat F)^\dagger \hat x_1 (\hat X(-m)\hat F) +\delta \notag\\
    \hat p_2'= (\hat X(-m)\hat F)^\dagger \hat p_1 (\hat X(-m)\hat F)+ \hat p_2.
\end{align}
If the input state is $\hat D_{\hat x}\ket{\phi}$, then the corresponding quantum circuit diagram is as follows
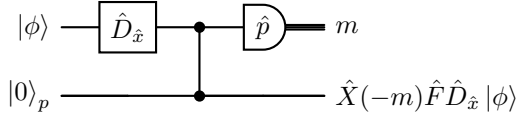
\begin{center}
    \begin{quantikz}
        \lstick{$\ket{\phi}$} & \gate{\hat D_{\hat x}} & \ctrl{1}  & \meterD{\hat p} & \cw  \rstick{$m$}\\
        \lstick{$\ket{0}_p$} & \qw & \control{}  &\qw & \qw 
        \rstick{$\hat X(-m)\hat F\hat D_{\hat x}\ket{\phi}$}\\
    \end{quantikz}   
\end{center}
If we assume that $\hat D_{\hat x}$ is a gate acting diagonally in the $\hat x$ eigenbasis, it can be commuted with the entangling gate which gives the following quantum circuit
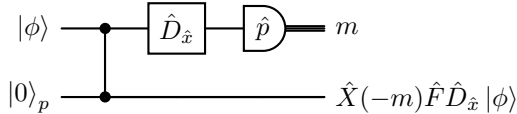
\begin{center}
    \begin{quantikz}
        \lstick{$\ket{\phi}$} & \ctrl{1} & \gate{\hat D_{\hat x}} & \meterD{\hat p} & \cw  \rstick{$m$}\\
        \lstick{$\ket{0}_p$} & \control{} & \qw &\qw & \qw 
        \rstick{$\hat X(-m)\hat F\hat D_{\hat x}\ket{\phi}$}\\
    \end{quantikz}   
\end{center}
Finally, applying $\hat D_{\hat x}$ followed by a measurement of $\hat p$ is equivalent to a measurement of $\hat D_{\hat x}^\dagger \hat p\hat D_{\hat x}$. This allows us to "apply" the gate $\hat D_{\hat x}$ by a suitable choice of the measurement basis as shown on the following quantum circuit

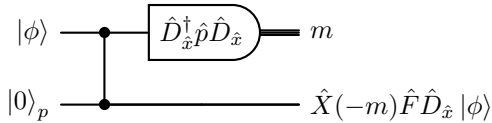
\begin{center}
    \begin{quantikz}
        \lstick{$\ket{\phi}$} & \ctrl{1} & \meterD{\hat D_{\hat x}^\dagger\hat p \hat D_{\hat x}} & \cw  \rstick{$m$}\\
        \lstick{$\ket{0}_p$} & \control{} & \qw & \qw 
        \rstick{$\hat X(-m)\hat F\hat D_{\hat x}\ket{\phi}$}\\
    \end{quantikz}   
\end{center}
Gaussian operations are easily implemented in this way, as when $\hat D_{\hat x}=e^{is\hat x}$, $\hat D_{\hat x}^\dagger\hat p \hat D_{\hat x}=\hat p+s$, which is equivalent to measuring $\hat p$ and adding $s$ to
the result. The case $\hat D_{\hat x}=e^{is\hat x^2/2}$, induces a measurement in $\hat D_{\hat x}^\dagger\hat p \hat D_{\hat x}=\hat p+s\hat x$, representing a measurement in a re-scaled rotated quadrature basis. For the gate $\hat D_{\hat x}=e^{is\hat x^3/3}$ (required for universal single-mode computations), the measurement is given in the basis of $\hat D_{\hat x}^\dagger\hat p \hat D_{\hat x}=\hat p+s\hat x^2$. To apply a sequence of gates $\hat D_1, \dots, \hat D_n$, we can consider the following diagram

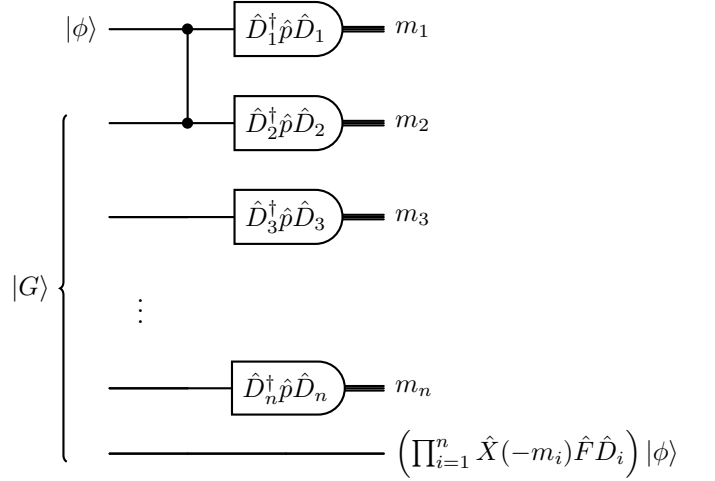
\begin{center}
    \begin{quantikz}
        \lstick{$\ket{\phi}$} & \ctrl{1} & \meterD{\hat D_1^\dagger\hat p \hat D_1} & \cw  \rstick{$m_1$}\\
        \lstick[5]{$\ket{G}$} & \control{} & \meterD{\hat D_2^\dagger\hat p \hat D_2} & \cw  \rstick{$m_2$}\\
        &\qw & \meterD{\hat D_3^\dagger\hat p \hat D_3} & \cw  \rstick{$m_3$}\\
        \push{~~~~~~~\vdots}\\
        &\qw & \meterD{\hat D_n^\dagger\hat p \hat D_n} & \cw  \rstick{$m_n$}\\
        & \qw &\qw & \qw         \rstick{$\left(\prod_{i=1}^n\hat X(-m_i)\hat F\hat D_i\right)\ket{\phi}$}\\
    \end{quantikz}   
\end{center}
Here, $\ket{G}$ is the linear graph state, associated with a $1$D chain graph. Now, we aim to understand how the noise evolves after each step. Let us write the quadrature operators of the initial state $\hat x_1 +\delta_x$ and $\hat p_1+\delta_p$, where the terms $\delta_x$ and $\delta_p$ correspond to the noise term created by earlier computation. We want to comprehend how these terms evolve as we perform the gate teleportation scheme.\\

{$\blacktriangleright$ \bf The $e^{is\hat x} $ gate}\\ 
This scenario corresponds to the following quantum circuit diagram:
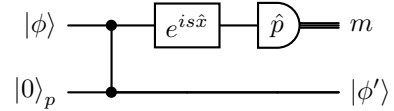
\begin{center}
    \begin{quantikz}
        \lstick{$\ket{\phi}$} & \ctrl{1} & \gate{e^{is\hat x}} & \meterD{\hat p} & \cw  \rstick{$m$}\\
        \lstick{$\ket{0}_p$} & \control{} & \qw &\qw & \qw 
        \rstick{$\ket{\phi'}$}\\
    \end{quantikz}   
\end{center}
Following a similar computation as before, we find that the final quadrature operators are expressed as (where we denote $\delta_m$ as the measurement noise and $\delta_g$ as the noise coming from imperfect squeezing)
\begin{align}
    \hat x_2=m-\hat p_1-s -\delta_p +\delta_m && \hat p_2=\hat x_1+\delta_x+\delta_g.
\end{align}
Thus, we observe that
\begin{align}
    \delta_x'=\delta_p+\delta_m && \delta_p'=\delta_x+\delta_g.
\end{align}
This indicates that the noise terms are exchanged between the quadratures (due to the effect of $\hat F$) and are amplified by both the measurement noise and the squeezing noise.

{$\blacktriangleright$ \bf The $e^{is\hat x^2/2} $ gate}\\ 
The specific quantum operation involving the $e^{is\hat x^2/2}$ gate is represented by the following diagram
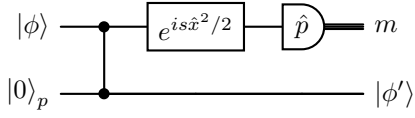
\begin{center}
    \begin{quantikz}
        \lstick{$\ket{\phi}$} & \ctrl{1} & \gate{e^{is\hat x^2/2}} & \meterD{\hat p} & \cw  \rstick{$m$}\\
        \lstick{$\ket{0}_p$} & \control{} & \qw &\qw & \qw 
        \rstick{$\ket{\phi'}$}\\
    \end{quantikz}   
\end{center}
In this context, the corresponding equations for the quadrature operators are given by:
\begin{align}
    &\hat x_2=m-\hat p_1-s\hat x_1 -\delta_p +\delta_m -s\delta_x\notag\\
    &\hat p_2=\hat x_1+\delta_x+\delta_g.
\end{align}
Consequently, the relationships between the noise terms are expressed as:
\begin{align}
    \delta_x'=\delta_p+\delta_m+ s\delta_x && \delta_p'=\delta_x+\delta_g.
\end{align}
It is noteworthy that in this scenario, noise in the $x$ quadrature arises from the contribution of three terms. A significant distinction from the previous situation lies in the presence of the noise term $s\delta_x$, which is directly proportional to the evolution parameter. By confining our considerations to Gaussian gates, we observe that the noise terms are expressed as the sum of random terms. Drawing upon concepts developed in the section on quantum teleportation, we are equipped to establish bounds on the probability that this noise exceeds a certain threshold. The detailed argument are exposed in the next section (\ref{sec: dealing with the noise})\\

{$\blacktriangleright$ \bf The $e^{is\hat x^3/3} $ gate}\\
This scenario corresponds to the quantum circuit depicted below
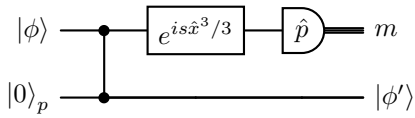
\begin{center}
    \begin{quantikz}
        \lstick{$\ket{\phi}$} & \ctrl{1} & \gate{e^{is\hat x^3/3}} & \meterD{\hat p} & \cw  \rstick{$m$}\\
        \lstick{$\ket{0}_p$} & \control{} & \qw &\qw & \qw 
        \rstick{$\ket{\phi'}$}\\
    \end{quantikz}   
\end{center}
Similarly, we obtain the following expressions:
\begin{align}
    \hat x_2&=m-\hat p_1-s\hat x_1^2 -2s\hat x_1\delta_x-s\delta_x^2-\delta_p +\delta_m \notag\\
    \hat p_2&=\hat x_1+\delta_x+\delta_g.
\end{align}
Consequently, we observe that
\begin{align}
    \delta_x'=\delta_p+\delta_m+ s\delta_x^2+2s\hat x_1\delta_x && \delta_p'=\delta_x+\delta_g.
\end{align}
In this instance, two intricate terms contribute to the new noise. The quadratic term $s\delta_x^2$ proves challenging to control. The other term, $2s\hat x\delta_x$, exacerbates the situation as the noise now depends on the operator's value. Thus, the noise becomes unmanageable when employing the cubic gate. To conduct Measurement-Based Quantum Computation (MBQC) in a noisy environment, one must circumvent the use of these gates. However, they are indispensable for achieving universal quantum computation. Consequently, alternative strategies must be employed to navigate this challenge, such as the use of magic states. However, as this is not the aim of this article, we won't delve into the different ideas that can be used.

\subsection{Dealing with the noise}\label{sec: dealing with the noise}
In our exploration of Measurement-Based Quantum Computing (MBQC), we observed that the output states of various protocols are characterized by quadrature operators, defined by equations of the form
\begin{equation}
    \hat x=\hat x_\text{ideal}+\delta_\text{noise},
\end{equation}
and similar for $\hat p$. The noise term $\delta_\text{noise}$, arises from two distinct contributions
\begin{itemize}
    \item {\bf Noise from Imperfect Measurement Devices:} This noise can be modeled as a sum of independent Gaussian random variables, each corresponding to a specific measurement. The overall noise from these contributions can then be effectively modeled by a single Gaussian term, with its variance being the sum of the variances of all individual contributions.
    \item {\bf Noise from Imperfectly Prepared Graph State:} To bound the noise originating from a single nullifier (as per Proposition \ref{pro: concentration on g}), we recognize that assuming independence between multiple nullifiers may not be valid. Nevertheless, we can derive the following inequality
    \begin{subequations}
        \begin{align}
            \P(\abs{\hat g_1+\hat g_2}\geq \delta)&\leq \P(\abs{\hat g_1}\geq \delta/2 \text{ or } \abs{\hat g_2}\geq \delta/2)\notag\\
            &\leq \P(\abs{\hat g_1} \leq \delta/2)+\P(\abs{\hat g_2}\geq \delta/2)\notag\\
            &\leq \eta+e^{-\delta^2/4\Delta^2}+\eta+e^{-\delta^2/4\Delta^2}\notag\\
            &=2\eta+2e^{-\delta^2/4\Delta^2}.
        \end{align}
    \end{subequations}
    This leads to the inequality
    \begin{equation}
        \P(\abs{\hat g_1+\hat g_2}\leq \delta)\geq 1-2\eta-2e^{-\delta^2/4\Delta^2}.
    \end{equation}
\end{itemize}
By combining this concentration inequality with those from proposition \ref{pro: concentration g delta}, we obtain a bound on the cumulative noise magnitude.

\subsection{Metrology}
For our final exploration of applications, we turn our attention to metrology. Discussing the implications of the inequality $O^\Delta \geq 1-\eta$ in terms of measurement precision is crucial when integrating graph states from the state certification protocol into a metrology protocol. In this context, we focus on the simplest scenario, considering a trivial graph state composed of a single mode. This state's physical approximation is a finitely squeezed single-mode state represented by

\begin{equation}
    \ket{\psi_\sigma}=\frac{1}{(2\pi\sigma^2)^{1/4}}\int dx e^{-\frac{x^2}{4\sigma^2}} \ket{x}.
\end{equation}
Such that $\operatorname{Var}_{\ket{\psi_\sigma}}(\hat x)= \sigma^2$ and $\operatorname{Var}_{\ket{\psi_\sigma}}(\hat p)= \frac{1}{4\sigma^2}$. These states prove effective for measuring momentum translations generated by $\hat x$. For such evolutions, the quantum Fisher information simplifies to $\mathcal Q=4\operatorname{Var}{\ket{\psi\sigma}}(\hat x) =4 \sigma^2$. Although we anticipate that states emerging from the certification protocol are good at momentum measurements, we lack direct access to their variance in $\hat x$, given only our knowledge of the overlap. Despite this limitation, we can still make a crude reasoning, that will give a lower bound on $\mathcal Q$ and that shows that the measurement precision can be high. Drawing on the earlier bound regarding the noisy measurement of nullifiers (prop. \ref{pro: concentration g delta}), we know that
\begin{equation}
    \P[\abs{\hat p+\delta}\leq p^*]\geq 1-\eta-\theta,
\end{equation}
where we defined $p^*=(\mu+\Delta)\sqrt{\ln(\frac{2}{\theta})}$, for any $0<\theta\leq 1$. We now consider a metrology protocol running as follows. One measures the operators $\hat p$. Whenever the measurement outcome is greater than $p^*$, the result is discarded, and otherwise it is kept. The variance of the kept measurement outcomes is then necessarily less than $(p^*)^2$, meaning that the associated Fisher information has to be at least $1/(p^*)^2=\frac{1}{(\mu+\Delta)^2\ln(2/\theta)}$. Finally, since a fraction $\eta+\theta$ of measurement rounds is wasted, the final quantum Fisher information is greater than
\begin{equation}
    \mathcal{Q}\geq \frac{(1-\eta-\theta)}{(\mu+\Delta)^2\ln(2/\theta)}.
\end{equation}
By optimizing over $\theta$, we see that the lower bound on the quantum Fisher information can be high, whenever $\eta$, $\mu$, and $\Delta$ are small enough.

\section{Comparison to other works} \label{Sec Comparison}

In this study, we have adapted the protocol initially formulated in \cite{takeuchi_resource-efficient_2019} to the continuous variables setting to deal with the inevitable difficulties of physical states and measurements in CV. The performance in terms of resources essentially remains unchanged. 
While the \cite{takeuchi_resource-efficient_2019} discusses comparisons with existing literature, it emphasizes the challenge of making a fair comparison due to differences in formulation. 
Our formulation of the different soundness statement allows us to make a more straightforward comparison with other works in the literature \cite{aolita_reliable_2015,chabaud_efficient_2021,wu_efficient_2021}.
In all cases, our protocol improves on previous results, either in scaling directly, or in removing the assumption of identical independent distribution (i.i.d.) of the source, or both. In many ways, this improvement is not surprising since our protocol is tailored to CV graph states, a highly specialized set of states most amenable to certification locally.

We focus on the soundness statement (thm. \ref{thm: soundness}), utilizing the parameter choices outlined in section \ref{sec: special cases} to simplify equation (\ref{eq: soundness non-conditionnal}), yielding
\begin{align}
    \Tr(\hat \Pi_\text{acc}\otimes[\1-\hat\Pi_1\times\cdots\times\hat \Pi_k]\hat \rho)\leq\frac{1}{\lambda}\left(\frac{4nk}{1-\frac{k}{\lambda^2}}+1\right).
\end{align}
From this, it is evident that achieving $\Tr(\hat \Pi_\text{acc}\otimes[\1-\hat\Pi_1\times\cdots\times\hat \Pi_k]\hat \rho)\leq \eta$ requires a total number of registers on the order of
\begin{equation}
    N_\text{tot}=2nN_\text{test}=\mathcal O\left(\frac{n^3k^2}{\eta^2}\ln(\frac{nk}{\eta})\right).
\end{equation}
This scaling may appear more efficient than the original formulation \cite{takeuchi_resource-efficient_2019}, which was $N_\text{tot}=\mathcal O\left(n^5\ln(n)\right)$, but it was obtained by leaving $n$ the only free parameter, which means that they were forcing $n=1/\eta$.  Our scaling recovers theirs in a more general manner, allowing flexibility in adjusting the bound tightness (small $\eta$) while keeping $n$ fixed. Furthermore, we provide the dependency on the number of copies $k$ retained at the protocol's conclusion, enabling comparison with other protocols obtained in the literature. It is then important to notice, that although we take into consideration noisy measurement and imperfect state, the scaling is not worsened compared to the protocol exposed in \cite{takeuchi_resource-efficient_2019}.

In the work of Ya-Dong et al \cite{wu_efficient_2021}, they devised a protocol for verifying various families of continuous-variable states, employing a similar formulation for soundness
\begin{equation}\label{eq: soundness wu}
    \Tr(\hat T(\1-\ketbra{\psi}^{\otimes k})\hat \rho)\leq \eta,
\end{equation}
which can be interpreted as a boundary on the probability of falsely accepting a state. Here
$\hat T$ is the POVM associated with whether the state is accepted or not by the protocol, $\ketbra{\psi}$ denotes the ideal target state, $n$ signifies the size of the state $\ket{\psi}$ and $k$ stands for the number of copies kept at the end of the protocol. They showed that to achieve a probability lower than $\eta$, the number of registers (copies of the states) must have the following asymptotic limit
\begin{align}
    N_\text{tot}=\mathcal{O}\left(\frac{n^7k^4}{\eta^6}\operatorname{Poly}\left(\ln\frac{nk}{\eta}\right)\right).
\end{align}

Comparing their formulation to ours by establishing the correspondence $\hat T\leftrightarrow \hat \Pi_\text{acc}$ and $\ketbra{\psi}^k\leftrightarrow \hat\Pi_1\times\cdots\times\Pi_k$, we observe that our approach is considerably more efficient. This discrepancy makes sense as their method enables the certification of a broader class of states than just the CV graph states. It's noteworthy that our protocol attains the efficiency of known protocols under the i.i.d (identically and independently distributed copies) hypothesis, even though we do not assume i.i.d. \cite{wu_efficient_2021,chabaud_efficient_2021}
\begin{equation}
    N_\text{tot}=\mathcal{O}\left(\frac{\operatorname{Poly}\left(\ln\frac{1}{\eta}\right)}{\eta^2}\right).
\end{equation}

Another work worth comparing is that of Chabaud et al \cite{chabaud_efficient_2021}. In this paper, the authors approach the problem of state certification slightly differently. They attempt to estimate the fidelity of a given $k$-system state $\hat \rho^k$ with some target state $\ket{C}^{\otimes k}$, where $\ket{C}$ has finite support over the Fock basis. They introduced a protocol such that for $\epsilon>0$, the fidelity estimate $F_C(\hat \rho)^k$ obtained is $\epsilon$-close to the `true' fidelity $F(\ket{C}^{\otimes k},\hat \rho^{ k})$ whenever the total number of copies $N_\text{tot}$ satisfies
\begin{equation}
    N_\text{tot}=\mathcal O\left(\frac{k^{7+\frac{4c}{p}}}{\epsilon^{4+\frac{4c}{p}}}\right),
\end{equation}
or the protocol abort with probability $P=\mathcal O\left(\frac{1}{\operatorname{Poly}(k,1/\epsilon)}\right)$. Here, $N_\text{tot}$ represents the total number of registers, $k$ denotes the number of registers left after the measurement step, $p$ is a free integer parameter specifying how the protocol should run, and $c$ serves as an upper-bound of the support of $\ket{C}$ over the Fock basis: $\ket{C}=\sum\limits_{n=0}^{c-1}c_n\ket{n}$.

By setting a large $p\in\mathbb{N}^\ast$, the scaling can be made arbitrarily close to $\mathcal O\left(\frac{k^7}{\epsilon^4}\right)$. While the comparison to our work is less straightforward, as perfect graph states cannot be expressed with a finite number of Fock states, and $\epsilon$ cannot be directly related to our value of $\eta$, it is theoretically possible to reframe this formulation into one closer to ours. In such a scenario, our scaling would outperform that obtained by \cite{chabaud_efficient_2021}. This is unsurprising, considering that the protocol is adapted to a different and larger class of states than the graph states. Furthermore, the main result obtained by the authors is formulated under the i.i.d hypothesis, which is unnecessary in our work. The extension to the general case is presented in the appendices of \cite{chabaud_efficient_2021}.

Lastly, let us discuss the comparison with \cite{aolita_reliable_2015}. In this paper, the authors devise a protocol to verify different classes of continuous-variable states. For verifying Gaussian states, they claim, using the i.i.d hypothesis, that their protocol requires 
\begin{equation}
    N_\text{tot}=\mathcal O\left(\frac{s_\text{max}^4\sigma_2^2n^7}{\epsilon^2\ln(\frac{1}{1-\alpha})}\right)
\end{equation}
copies of the targeted states. Here, $s_\text{max}$ and $\sigma_2$ are parameters characterizing the squeezing of both the produced and targeted states, $\alpha$ is a failure probability, $n$ is the number of modes of the targeted state, and $\epsilon$ is the error in the fidelity estimation. Once again, the comparison is not straightforward, as graph states are Gaussian states that require infinite squeezing. However, by introducing finite values for $s_\text{max}$ and $\sigma_2$, one would obtain a certification protocol closely approximating the perfect graph state, which could then be related to a genuine protocol for verifying graph states. Ignoring the overhead required to make this translation, we observe that the scaling presented in the paper is similar to ours. However, our results do not use the i.i.d hypothesis, making our protocol more efficient for the certification of CV graph states. This is unsurprising since the protocol is devised for the very specific class of Gaussian states, allowing for the exploitation of specific properties to create an efficient certification protocol.

\section{Conclusion}
In this work, we have successfully adapted and expended the protocol originally introduced in \cite{takeuchi_resource-efficient_2019} to the continuous variables regime. The inherent physical constraints of CV have been addressed, such as the impossibility of producing infinite energy states or measuring with perfect precision. We have provided alternative ways of formulating the bounds so that it was more feasible to compare with existing work. Lastly, we have discussed how states certified by our protocol can be used in further application, thereby certifying said applications and addressing the problem of how imperfections alter the results. This work is another step in the direction of expanding the usability of continuous variable systems in concrete applications.

\begin{acknowledgments}
   The authors thank Frédéric Grosshans for fruitful discussions and his expertise on continuous variables systems. They are also grateful to Ulysse Chabaud for the interesting discussions while reviewing the work and discussing about comparison with the literature. D.M. acknowledges funding from
Plan France 2030 through the ANR-22-PETQ-0006
NISQ2LSQ project, as well as Horizon Europe Research and Innovation Actions under Grant Agreement no. 101080173 (CLUSTEC).
\end{acknowledgments}

\bibliographystyle{unsrturl} 
\bibliography{refs}

\appendix
\onecolumngrid

\section{Lemmas}
In this section, we formulate and prove various lemmas that are needed for the proof of the multiple theorem and proposition of this paper. The first one is a result formulated by Serfling \cite{serfling_probability_1974} that is the backbone of how the protocol certifies that the state at the end is close to the desired one. 

\begin{Lem}\label{lemme: commutation exp}
    For two operators $A$ and $B$ satisfying $[A,[A,B]]=0$, we have the relation:
    \begin{equation}
        e^ABe^{-A}=B+[A,B]
    \end{equation}
\end{Lem}

\begin{preuve}
We introduce two functions of the real variable $t$:
    \begin{align}
        \varphi(t)=e^{tA}Be^{-tA} && \psi(t)=B+t[A,B]
    \end{align}
    We can compute the derivatives of both functions:
    \begin{equation}
        \varphi'(t)=Ae^{tA}Be^{-tA}-e^{tA}Be^{-tA}A=A\varphi(t)-\varphi(t)A=[A,\varphi(t)]
    \end{equation}
    and
    \begin{equation}
        \psi'(t)=[A,B]=[A,B+t[A,B]]=[A,\psi(t)]
    \end{equation}
    where we used here the hypothesis on the commutation between $A$ and $[A,B]$. We thus see that both function $\psi$ and $\varphi$ satisfy the same first-order linear differential equation. Since moreover $\psi(0)=\varphi(0)=B$, the Cauchy-Lipschitz theorem allows us to conclude that both functions are the same. 
\end{preuve}

\begin{Lem}\label{lemme: serfling}
    We consider $N$ (not necessarily distinct) values: $x_1,\dots,x_N$. For a random variable $\Pi=(I_1,\dots,I_n)$ which correspond to a uniform sampling of $n$ indices without replacement (\textit{i.e.} the $I_j$ takes distinct values between $1$ and $N$), define the associated random variables $X_j=x_{I_j}$ as well as their sum  $S_n=\sum\limits_{j=1}^n X_j$. Then for $t>0$,
    \begin{equation}
        \P\left[\frac{1}{n}S_n\geq \mu +t\right]\leq \exp[-\frac{2nt^2}{(1-f_n^\ast)(b-a)^2}],
    \end{equation}
    with $f_n^\ast=\frac{n-1}{N}$, $a=\min\limits_{1\leq i\leq N}x_i$, $b=\max\limits_{1\leq i\leq N}x_i$, $\mu=\frac{1}{N}\sum\limits_{i=1}^N x_i$.
\end{Lem}
\begin{preuve}
    We don't provide proof here, see the paper written by Serfling \cite{serfling_probability_1974}
\end{preuve}

The lemma formulated by Serfling is only on probability lemma. Its relation to quantum state verification and quantum measurement is not obvious. Here we give a result that makes the link. It has originally been formulated and demonstrated in \cite{tomamichel_largely_2017}

\begin{Lem}
    Consider a set of binary random variables $Z = (Z_1, Z_2, \dots , Z_m)$ with $Z_i$ taking values in $\{0, 1\}$ and $m = n + k$. Let $\Pi$ be an independent, uniformly distributed random variable corresponding to a uniform sampling without replacement of $k$ indices between $1$ and $m$ \textit{i.e.} $\Pi\subset \{1,\dots,m\}$ and $\#\Pi=k$. Then for $\nu>0$,
    \begin{equation}
        \P\left[\frac{1}{n}\sum_{i\in\bar{\Pi}}Z_i\geq \frac{1}{k}\sum_{i\in \Pi}Z_i+\nu\right ]\leq \exp[-2\nu^2\frac{nk^2}{(n+k)(k+1)}],
    \end{equation}
    where $\bar{\Pi}$ denotes the complementary of $\Pi$. Interestingly, this bound only asks the independence between $\Pi$ and the $Z_i$'s. There is no assumption on the distributions of the $Z_i$'s and they can be correlated.
\end{Lem}
\begin{preuve}
    For completeness, we redo the proof, by mostly copying the argument of \cite{tomamichel_largely_2017}.     We can show this using Serfling's statement (Lemma \ref{lemme: serfling}). For $z\in\{0,1\}^m$, we define $\mu(z)=\frac{1}{m}\sum_{i=1}^m z_i$. And we write:
    \begin{subequations}
        \begin{align}
            \P\left[\frac{1}{n}\sum_{i\in\bar{\Pi}}Z_i\geq \frac{1}{k}\sum_{i\in \Pi}Z_i+\nu\right ]&=\sum_{z\in\{0,1\}^m}\P\left[Z=z ~\&~ \frac{1}{n}\sum_{i\in\bar{\Pi}}z_i\geq \frac{1}{k}\sum_{i\in \Pi}z_i+\nu\right] ~~ (\text{using the law of total probability})\\
            &=\sum_{z\in\{0,1\}^m}\P\left[Z=z\right]\P\left[ \frac{1}{n}\sum_{i\in\bar{\Pi}}z_i\geq \frac{1}{k}\sum_{i\in \Pi}z_i+\nu\right] ~~ (\text{by independence between $Z$ and $\Pi$})\\
            &=\sum_{z\in\{0,1\}^m}\P\left[Z=z\right]\P\left[ \frac{1}{n}\sum_{i\in\bar{\Pi}}z_i\geq \frac{1}{k}\left(m\mu(z)-\sum_{i\in \bar{\Pi}}z_i\right)+\nu\right] \\
            &=\sum_{z\in\{0,1\}^m}\P\left[Z=z\right]\P\left[ \left(\frac{1}{n}+\frac{1}{k}\right)\sum_{i\in\bar{\Pi}}z_i\geq \frac{m}{k}\mu(z)+\nu\right] \\
            &=\sum_{z\in\{0,1\}^m}\P\left[Z=z\right]\P\left[ \frac{n+k}{nk}\sum_{i\in\bar{\Pi}}z_i\geq \frac{m}{k}\mu(z)+\nu\right] \\
            &=\sum_{z\in\{0,1\}^m}\P\left[Z=z\right]\P\left[ \frac{m}{nk}\sum_{i\in\bar{\Pi}}z_i\geq \frac{m}{k}\mu(z)+\nu\right] \\
            &=\sum_{z\in\{0,1\}^m}\P\left[Z=z\right]\P\left[ \frac{1}{n}\sum_{i\in\bar{\Pi}}z_i\geq \mu(z)+\frac{k\nu}{m}\right].
        \end{align}
    \end{subequations}
    At this point, we are exactly in the form of Serfling's inequality, with $t=k\nu/m$, $a=0$, $b=1$ $\mu=\mu(z)$ and $f_n^\ast=(n-1)/m$. Thus independently of the value of $z$, we have:
    \begin{equation}
        \P\left[ \frac{1}{n}\sum_{i\in\bar{\Pi}}z_i\geq \mu(z)+\frac{k\nu}{m}\right]\leq \exp[-2n\left(\frac{k\nu}{m}\right)^2\frac{1}{1-\frac{n-1}{m}}]=\exp[-2\nu^2\frac{nk^2}{(k+n)(k+1)}].
    \end{equation}
    Since this bound does not depend on $z$, we can simply sum the $\P(Z=z)$ and get the result:
    \begin{equation}
        \P\left[\frac{1}{n}\sum_{i\in\bar{\Pi}}Z_i\geq \frac{1}{k}\sum_{i\in \Pi}Z_i+\nu\right ]\leq \exp[-2\nu^2\frac{nk^2}{(n+k)(k+1)}].
    \end{equation}
\end{preuve}
 We now give the formulation of two lemmas, the second one being a generalization of the first one. So strictly speaking, the first one is contained in the second, but we still introduce both for clarity. It is the technical lemmas, that allow us to go from global information about the collection of unmeasured states to specific information on the states that are kept at the end of the protocol. 

\begin{Lem}\label{lemma: lNn}
     We take $n,N\in \N$. For real numbers $\lambda_1,\dots,\lambda_n$ lets denote:
     \begin{equation}
         L_N^{n}(\lambda_1,\dots,\lambda_n)=\sum\limits_{\substack{i_1,\dots,i_n\in\{0,1\}\\i_1+\cdots+i_n\geq N}} \prod_{k=1}^n \lambda_k^{i_k}(1-\lambda_k)^{1-i_k},
     \end{equation}
     with the same convention as before: any real number to the power of $0$ is equal to $1$.\\
     If $\forall 1\leq k\leq n,~ \lambda_k\in [0,1]$ then:
     \begin{equation}
         N L_N^n(\lambda_1,\dots,\lambda_n)\leq \lambda_1+\cdots+\lambda_n.
     \end{equation}
 \end{Lem}
 \begin{preuve}
     We show it by induction on $n$.\\
     First if $n=1$, lets take $N\in\N$. If $N>1$ then $L_N^1(\lambda_1)=0$ and the inequality is trivial. For $N=0$ this is also easy to see. Finally if we take $N=1$, we have:
     \begin{equation}
         1\times L_1^1(\lambda_1)=\sum_{i_1\in\{0,1\},~i_1\geq 1 }\lambda_1^{i_1}(1-\lambda_1)^{1-i_1}=\lambda_1.
     \end{equation}

     We consider a general $n\in \N$ and lets assume that the result is true for $n$. We take $N\in \N$ and $\lambda_1,\dots,\lambda_n\in[0,1]$. If we consider the function:
     \begin{equation}
         \lambda_{n+1}\mapsto \lambda_1+\cdots+\lambda_ {n+1}-NL^{n+1}_N(\lambda_1,\dots,\lambda_{n+1}),
     \end{equation}
     we see that it is an affine function in $\lambda_{n+1}$. Indeed when we take a look at the expression of $L_N^{n+1}(\lambda_1,\dots,\lambda_{n+1})$ we are taking the sum of products of $n$ factors. There is exactly one factor associated with the variable $\lambda_k$, which is either $\lambda_k$ or $1-\lambda_k$ depending on the value of $i_k$. This means that the power of $\lambda_{n+1}$ appearing in each product is at most one. Thus the map above is indeed an affine function of $\lambda_{n+1}$.\\
     
     Since this affine map is defined on $[0,1]$, we know that when keeping all other values constant, the extreme values will be reached on the boundary. This means that we only have to verify the inequality when $\lambda_{n+1}=0\text{ or }1$. Lets first take $\lambda_{n+1}=0$. We compute:
     \begin{subequations}
        \begin{align}
            NL_N^{n+1}(\lambda_1,\dots,\lambda_n,0)&=N\sum_{\substack{i_1,\dots,i_{n+1}\in\{0,1\}\\i_1+\cdots+i_{n+1}\geq N}} \prod_{k=1}^{n+1} \lambda_k^{i_k}(1-\lambda_k)^{1-i_k}\\
            &=N\sum_{\substack{i_1,\dots,i_{n+1}\in\{0,1\}\\i_1+\cdots+i_{n+1}\geq N}} \left(\prod_{k=1}^n \lambda_k^{i_k}(1-\lambda_k)^{1-i_k}\right)\underbrace{0^{i_{n+1}}(1-0)^{1-i_{n+1}}}_{\substack{=0\text{ if } i_{n+1}=1\\ =1 \text{ if }i_{n+1}=0}}\\
            &=N\sum_{\substack{i_1,\dots,i_n\in\{0,1\}\\i_1+\cdots+i_n\geq N}} \prod_{k=1}^n \lambda_k^{i_k}(1-\lambda_k)^{1-i_k}\\
            &=NL_N^n(\lambda_1,\dots,\lambda_n)\\
            &\leq \lambda_1+\cdots+\lambda_n\text{ (thanks to the induction hypothesis)}\\
            &=\lambda_1+\cdots+\lambda_{n+1}.            
        \end{align}         
     \end{subequations}
    And similarly, if $\lambda_{n+1}=1$ we have:
    \begin{subequations}
        \begin{align}
            NL_N^{n+1}(\lambda_1,\dots,\lambda_n,1)&=N\sum_{\substack{i_1,\dots,i_{n+1}\in\{0,1\}\\i_1+\cdots+i_{n+1}\geq N}} \prod_{k=1}^{n+1} \lambda_k^{i_k}(1-\lambda_k)^{1-i_k}\\
            &=N\sum_{\substack{i_1,\dots,i_{n+1}\in\{0,1\}\\i_1+\cdots+i_{n+1}\geq N}} \left(\prod_{k=1}^n \lambda_k^{i_k}(1-\lambda_k)^{1-i_k}\right)\underbrace{1^{i_{n+1}}(1-1)^{1-i_{n+1}}}_{\substack{=1\text{ if } i_{n+1}=1\\ =0 \text{ if }i_{n+1}=0}}\\
            &=N\sum_{\substack{i_1,\dots,i_n\in\{0,1\}\\i_1+\cdots+i_n+1\geq N}} \prod_{k=1}^n \lambda_k^{i_k}(1-\lambda_k)^{1-i_k}\\
            &=NL_{N-1}^n(\lambda_1,\dots,\lambda_n).
        \end{align}
    \end{subequations}
    To continue the proof we need to show:
    \begin{equation}
        L_N^n(\lambda_1,\dots,\lambda_n)\leq 1
    \end{equation}
    Indeed since we are only summing non-negative terms (this is because $\lambda_k\in[0,1]$), we have:
    \begin{subequations}
        \begin{align}
            L_N^n(\lambda_1,\dots,\lambda_n)&=\sum\limits_{\substack{i_1,\dots,i_n\in\{0,1\}\\i_1+\cdots+i_n\geq N}} \prod_{k=1}^n \lambda_k^{i_k}(1-\lambda_k)^{1-i_k}\\
            &\leq \sum\limits_{\substack{i_1,\dots,i_n\in\{0,1\}\\i_1+\cdots+i_n\geq 0}} \prod_{k=1}^n \lambda_k^{i_k}(1-\lambda_k)^{1-i_k}\\
            &=\prod_{k=1}^{n} \Big[\lambda_k +(1-\lambda_k)\Big]\\
            &=\prod_{k=1}^n 1=1.
        \end{align}
    \end{subequations}
    With this we can go back to the case $\lambda_{n+1}=1$:
    \begin{subequations}
         \begin{align}
            NL_N^{n+1}(\lambda_1,\dots,\lambda_n,1)&\leq NL^n_{N-1}(\lambda_1,\dots,\lambda_n)\\
            &= (N-1)L^n_{N-1}(\lambda_1,\dots,\lambda_n)+L^n_{N-1}(\lambda_1,\dots,\lambda_n)\\
            &\leq (N-1)L^n_{N-1}(\lambda_1,\dots,\lambda_n)+1\\
            &\leq \lambda_1+\cdots+\lambda_n+1\\
            &=\lambda_1+\cdots+\lambda_{n+1}
        \end{align}   
    \end{subequations}
    This finishes the proof of the lemma.
 \end{preuve}

\begin{Lem}\label{lemma: lNngen}
     We take $k,N\in \N$ and $n\in \N^*$. For real numbers $\lambda_1,\dots,\lambda_n$ lets denote:
     \begin{equation}
         L_N^{n}(\lambda_1,\dots,\lambda_n)=\sum\limits_{\substack{i_1,\dots,i_n\in\{0,1\}\\i_1+\cdots+i_n\geq N}} \prod_{k=1}^n \lambda_k^{i_k}(1-\lambda_k)^{1-i_k},
     \end{equation}
     with the same convention as before: any real number to the power of $0$ is equal to $1$. If $\forall 1\leq k\leq n,~ \lambda_k\in [0,1]$ then:
     \begin{equation}
         \binom{N}{k} L_N^n(\lambda_1,\dots,\lambda_n)\leq \sum_{1\leq a_1<\cdots<a_k\leq n} \lambda_{a_1}\times\cdots\times\lambda_{a_k},
     \end{equation}
     where the right term correspond to the sum of all $k$-fold product of the $\lambda_i$'s.
 \end{Lem}
 \begin{preuve}
     We show it by nested induction. We start with an induction over $k$ (its inductive steps will require an induction over $n$).\\

     For $k=0$, we impose by convention that the right term is equal to $1$ (there is exactly one zero-fold product of the $\lambda_i$'s: the empty product, which is equal to one by convention). We must then show that for all $n\geq 1$ and $N\geq 0$, we have
     \begin{equation}
         L_N^n(\lambda_1,\dots,\lambda_n)\leq 1.
     \end{equation}
 Indeed since we are only summing non-negative terms (this is because $\lambda_k\in[0,1]$), we have:
    \begin{subequations}
        \begin{align}
            L_N^n(\lambda_1,\dots,\lambda_n)&=\sum\limits_{\substack{i_1,\dots,i_n\in\{0,1\}\\i_1+\cdots+i_n\geq N}} \prod_{k=1}^n \lambda_k^{i_k}(1-\lambda_k)^{1-i_k}\\
            &\leq \sum\limits_{\substack{i_1,\dots,i_n\in\{0,1\}\\i_1+\cdots+i_n\geq 0}} \prod_{k=1}^n \lambda_k^{i_k}(1-\lambda_k)^{1-i_k}\\
            &=\prod_{k=1}^{n} \Big[\lambda_k +(1-\lambda_k)\Big]\\
            &=\prod_{k=1}^n 1=1.
        \end{align}
    \end{subequations}

    Now consider $k\geq 0$. We assume that the result is true for $k$ and want to show it for $k+1$. As stated before, we perform now an induction over $n$.\\
    For $n=1$, we aim to show that 
    \begin{equation}
        \binom{N}{k+1}L^1_N(\lambda_1)\leq \sum_{1\leq a_1<\cdots<a_{k+1}\leq 1} \lambda_{a_1}\times\cdots\times\lambda_{a_{k+1}}.
    \end{equation}
    Depending on the value of $N$ and $k$, we have
    \begin{align}
        L_N^1(\lambda_1)=\left\{\begin{matrix} 
            1 & \text{if }N=0\\
            \lambda_1 & \text{if }N=1\\
            0 &\text{if }N>1
        \end{matrix}\right. && \binom{N}{k+1}=\left\{\begin{matrix} 
            0 & \text{if }N=0\\
            1 & \text{if }N=1 \text{ and } k=0\\
            0 & \text{if }N=1 \text{ and } k>0\\
            \binom{N}{k+1} &\text{if }N>1
        \end{matrix}\right.
    \end{align}
    This means that the only way the left-hand side is non-zero (and thus the inequality is not trivially true) is when $N=1$ and $k=0$. In this situation, both the left and right-hand sides of the inequality we wish to show are equal to $\lambda_1$, which concludes the proof of the base case.\\
    
    Now consider the $n\geq 1$ and assume that the result is true for this value of $n$. For $N\in \N$ and $\lambda_1,\dots,\lambda_n\in[0,1]$, we consider the map
    \begin{equation}
         \lambda_{n+1}\mapsto \sum_{1\leq a_1<\cdots<a_{k+1}\leq n+1}\lambda_{a_1}\times\cdots\times\lambda_{a_{k+1}}-\binom{N}{k+1}L^{n+1}_N(\lambda_1,\dots,\lambda_{n+1}).
    \end{equation}
    We see that it is a affine function in $\lambda_{n+1}$.

     Indeed when we take a look at the expression of $L_N^{n+1}(\lambda_1,\dots,\lambda_{n+1})$ we are taking the sum of products of $n$ factors. There is exactly one factor associated with the variable $\lambda_k$, which is either $\lambda_k$ or $1-\lambda_k$ depending on the value of $i_k$. This means that the power of $\lambda_{n+1}$ appearing in each product is at most one. Similarly, the first sum adds the product of $k$ factors $\lambda_i$'s, all being different. Thus the map above is indeed an affine function of $\lambda_{n+1}$.\\
     
     Since this affine map is defined on $[0,1]$, we know that when keeping all other values constant, the extreme values will be reached on the boundary. This means that we only have to verify the inequality when $\lambda_{n+1}=0\text{ or }1$. Lets first take $\lambda_{n+1}=0$. We compute:
     \begin{subequations}
        \begin{align}
            \binom{N}{k+1}L_N^{n+1}(\lambda_1,\dots,\lambda_n,0)&=\binom{N}{k+1}\sum_{\substack{i_1,\dots,i_{n+1}\in\{0,1\}\\i_1+\cdots+i_{n+1}\geq N}} \prod_{l=1}^{n+1} \lambda_l^{i_l}(1-\lambda_l)^{1-i_l}\\
            &=\binom{N}{k+1}\sum_{\substack{i_1,\dots,i_{n+1}\in\{0,1\}\\i_1+\cdots+i_{n+1}\geq N}} \left(\prod_{l=1}^n \lambda_l^{i_l}(1-\lambda_l)^{1-i_l}\right)\underbrace{0^{i_{n+1}}(1-0)^{1-i_{n+1}}}_{\substack{=0\text{ if } i_{n+1}=1\\ =1 \text{ if }i_{n+1}=0}}\\
            &=\binom{N}{k+1}\sum_{\substack{i_1,\dots,i_n\in\{0,1\}\\i_1+\cdots+i_n\geq N}} \prod_{l=1}^n \lambda_l^{i_l}(1-\lambda_l)^{1-i_l}\\
            &=\binom{N}{k+1}L_N^n(\lambda_1,\dots,\lambda_n)\\
            &\leq \sum_{1\leq a_1<\cdots<a_{k+1}\leq n}\lambda_{a_1}\times\cdots\times\lambda_{a_{k+1}}\text{ (thanks to the induction hypothesis on $n$)}\\
            &=\sum_{1\leq a_1<\cdots<a_{k+1}\leq n+1}\lambda_{a_1}\times\cdots\times\lambda_{a_{k+1}}.           
        \end{align}         
     \end{subequations}
    And similarly, if $\lambda_{n+1}=1$ we have:
    \begin{subequations}
        \begin{align}
            \binom{N}{k+1}L_N^{n+1}(\lambda_1,\dots,\lambda_n,1)&=\binom{N}{k+1}\sum_{\substack{i_1,\dots,i_{n+1}\in\{0,1\}\\i_1+\cdots+i_{n+1}\geq N}} \prod_{l=1}^{n+1} \lambda_l^{i_l}(1-\lambda_l)^{1-i_l}\\
            &=\binom{N}{k+1}\sum_{\substack{i_1,\dots,i_{n+1}\in\{0,1\}\\i_1+\cdots+i_{n+1}\geq N}} \left(\prod_{l=1}^n \lambda_l^{i_l}(1-\lambda_l)^{1-i_l}\right)\underbrace{1^{i_{n+1}}(1-1)^{1-i_{n+1}}}_{\substack{=1\text{ if } i_{n+1}=1\\ =0 \text{ if }i_{n+1}=0}}\\
            &=\binom{N}{k+1}\sum_{\substack{i_1,\dots,i_n\in\{0,1\}\\i_1+\cdots+i_n+1\geq N}} \prod_{l=1}^n \lambda_l^{i_l}(1-\lambda_l)^{1-i_l}\\
            &=\binom{N}{k+1}L_{N-1}^n(\lambda_1,\dots,\lambda_n)\\
            &=\left[\binom{N-1}{k}+\binom{N-1}{k+1}\right]L_{N-1}^n(\lambda_1,\dots,\lambda_n)\text{ (By Pascal formula)}\\
            &\leq \sum_{1\leq a_1<\cdots<a_{k}\leq n}\lambda_{a_1}\times\cdots\times\lambda_{a_{k}}+\sum_{1\leq a_1<\cdots<a_{k+1}\leq n}\lambda_{a_1}\times\cdots\times\lambda_{a_{k+1}}
            \intertext{ (thanks to the induction hypothesis on $n$ and $k$)}
            &=\lambda_{n+1}\sum_{1\leq a_1<\cdots<a_{k}\leq n}\lambda_{a_1}\times\cdots\times\lambda_{a_{k}}+\sum_{1\leq a_1<\cdots<a_{k+1}\leq n}\lambda_{a_1}\times\cdots\times\lambda_{a_{k+1}}\\
            &=\sum_{1\leq a_1<\cdots<a_{k+1}\leq n+1}\lambda_{a_1}\times\cdots\times\lambda_{a_{k+1}}.
        \end{align}
    \end{subequations}
Since, when considering all the $k+1$ fold product of the $\lambda_i$'s, one can either take $\lambda_{n+1}$ in the product (corresponds to the first sum) or not (correspond to the second sum). This finishes the proof of the lemma.
 \end{preuve}
Finally, the last lemma is the extension of the two previous lemmas, but by replacing the real numbers $\lambda_i$ by POVMs.
\begin{Lem}\label{lemme: inequality operators}
    For $k$ and $N\in\N$, we have the following operator inequality
    \begin{equation}
        \binom{N}{k}\hat\Pi_\text{acc}^N\leq \sum_{1\leq a_1<\cdots<a_k\leq N_r} \hat \Pi_{a_1}\times\cdots\times\hat \Pi_{a_k}.
    \end{equation}
    
\end{Lem}
\begin{preuve}
    This is a direct consequence of the previous Lemma \ref{lemma: lNngen}. Both operators can be diagonalized in the same (generalized) basis $\displaystyle\Big\{\bigotimes_{i=1}^{N_r} \hat Z(\va{s}_i)\ket{G}\Big|\forall i,\va{s}_i\in \R^n\Big\}$. We thus check that the eigenvalues of the difference are indeed non-negative. We have seen the definition of the $\hat \Pi_i$ in term of the operators $\hat P_i^{(j)}$ with
    \begin{equation}
        \hat P^{(j)}=\frac{\epsilon}{\sqrt{\epsilon^2+\delta_j^2}} \int dg \exp(-\frac{g^2}{\epsilon^2+\delta_j^2}) \hat E_{g}^{(j)}.
    \end{equation}
    However we can check that $\hat E_{g}^{(j)}\hat Z(\va s)\ket{G}=\delta(g-s_j)\hat Z(\va s)\ket{G}$. Indeed we have, using commutation relations, that
    \begin{equation}
         \hat Z(\va s)^\dagger \hat g_i  \hat Z(\va s)=\hat g_i+s_i.
    \end{equation}
    And thus we get
    \begin{equation}
        \hat g_i  \hat Z(\va s)\ket{G}= \hat Z(\va s) \hat Z(\va s)^\dagger \hat g_i  \hat Z(\va s)\ket{G}= \hat Z(\va s)(\hat g_i+s_i)\ket{G}=s_i \hat Z(\va s)\ket{G}.
    \end{equation}
    And the formula $\hat E_{g}^{(j)}\hat Z(\va s)\ket{G}=\delta(g-s_j)\hat Z(\va s)\ket{G}$ stems from the fact that by definition, $\hat E_g^{(j)}$ is the projector over the eigenspace of $\hat g_j$ with the associated eigenvalue $g$. We can thus compute

    \begin{subequations}
        \begin{align}
            \hat P^{(j)}\hat Z(\va s)\ket{G}&=\frac{\epsilon}{\sqrt{\epsilon^2+\delta_j^2}}\int dg \exp(-\frac{g^2}{\epsilon^2+\delta_j^2})\hat E_g^{(i)}\hat Z(\va s)\ket{G}\\
            &=\frac{\epsilon}{\sqrt{\epsilon^2+\delta_j^2}}\int dg \exp(-\frac{g^2}{\epsilon^2+\delta_j^2})\delta(g-s_j)\hat Z(\va s)\ket{G}\\
            &=\frac{\epsilon}{\sqrt{\epsilon^2+\delta_j^2}} \exp(-\frac{s_j^2}{\epsilon^2+\delta_j^2})\hat Z(\va s)\ket{G}.
        \end{align}
    \end{subequations}
    And thus
    \begin{equation}\label{eq: eigenvalue equ pi}
        \hat \Pi \hat Z(\va s)\ket{G}=\left[\prod_{j=1}^n \frac{\epsilon}{\sqrt{\epsilon^2+\delta_j^2}} \exp(-\frac{s_j^2}{\epsilon^2+\delta_j^2})\right]\hat Z(\va s)\ket{G}.
    \end{equation}
    If now we put back the indices for the registers, we can define for $\va s_i\in \R^n$, $\lambda_i=\prod_{j=1}^n \frac{\epsilon}{\sqrt{\epsilon^2+\delta_j^2}} \exp(-\frac{(\va s_i)_j^2}{\epsilon^2+\delta_j^2})$ and get that
    \begin{equation}
        \hat \Pi_j\bigotimes_{i=1}^{N_r} \hat Z(\va{s}_i)\ket{G}=\lambda_j\bigotimes_{i=1}^{N_r} \hat Z(\va{s}_i)\ket{G}.
    \end{equation}
    Finally it is easy to check that $\lambda_i\in[0,1]$ and that
    \begin{align}
       &\left[\sum_{1\leq a_1<\cdots<a_k\leq N_r} \hat \Pi_{a_1}\times\cdots\times\hat \Pi_{a_k}-\binom{N}{k}\hat\Pi_\text{acc}^N\right]\bigotimes_{i=1}^{N_r} \hat Z(\va{s}_i)\ket{G}\notag\\
       &~~~~~~~~~~~~=\left[\sum_{1\leq a_1<\cdots<a_k\leq N_r}  \lambda_{a_1}\times\cdots\times \lambda_{a_k}-\binom{N}{k}L_N^{N_r}(\lambda_1,\dots,\lambda_{N_r})\right]\bigotimes_{i=1}^{N_r} \hat Z(\va{s}_i)\ket{G}.
    \end{align}
    Thus $\bigotimes_{i=1}^{N_r} \hat Z(\va{s}_i)\ket{G}$ is indeed an eigenvector of the difference of the two operators, with a positive eigenvalue, which can be deduced from the previous Lemma \ref{lemme: inequality operators}. This concludes the proof of the lemma.
\end{preuve}

\section{Theorems proofs}

\subsection{Proof of proposition \ref{pro: stab cv graph}}\label{proof: stab cv graph}
\begin{preuve}
We first show that:
\begin{equation}
    \hat g_a=\left(\prod_{\{i,j\}\in E} \hat{CZ}_{i,j}(\Omega_{i,j})\right)\hat p_a\left(\prod_{\{i,j\}\in E} \hat{CZ}_{i,j}(-\Omega_{i,j})\right)
\end{equation}
    To do so, we first consider only one edge $\{i,j\}\in E$. If $a\neq i,j$, then $\hat p_a$ and $CZ_{i,j}(\Omega_{i,j})$ act on different modes, and thus commute. And we have $CZ_{i,j}(\Omega_{i,j})\hat p_a CZ_{i,j}(-\Omega_{i,j})=\hat p_a$.\\
    If $a=i$ (it is symmetric if $a=j$), then we use the lemma \ref{lemme: commutation exp} for the operator $A=i\Omega_{i,j}\hat x_i\hat x_j$ and $B=\hat p_i$. We thus have:
    \begin{subequations}
       \begin{align}
        CZ_{i,j}(\Omega_{i,j})\hat p_i CZ_{i,j}(-\Omega_{i,j})&=e^ABe^{-A}=B+[A,B]\\
        &=\hat p_i +i\Omega_{i,j}\left[\hat x_i \hat x_j,\hat p_i\right]\\
        &=\hat p_i+i\Omega_{i,j}\hat x_j [\hat x_i,\hat p_i]\\
        &=\hat p_i-\Omega_{i,j}\hat x_j
        \end{align} 
    \end{subequations}
    Since the `x' operators commute among themselves, they also commute with the `controlled Z' operators (which are exponential of products of `x' operators). Thus when computing the full product, each time we consider a gate $CZ_{i,j}(\Omega_{i,j})$, either, $a\not\in \{i,j\}$ and it does not affect the operator or $a=i$, and we subtract to $\hat p_a$ $\Omega_{i,j}\hat x_j$. As expected, this leads to :
    \begin{equation}
        \left(\prod_{\{i,j\}\in E} \hat{CZ}_{i,j}(\Omega_{i,j})\right)\hat p_a\left(\prod_{\{i,j\}\in E} \hat{CZ}_{i,j}(-\Omega_{i,j})\right)=\hat p_a-\sum_{j\in N^{(a)}}\Omega_{a,j} x_j=\hat g_a
    \end{equation}
    Now, we denote $U=\prod_{\{i,j\}\in E} CZ_{i,j}(\Omega_{i,j})$, meaning that:
    \begin{equation}
        \hat g_i\ket{G}=U\hat p_i U^\dagger U\bigotimes_{j=1}^n\ket{0}^p_j=U\hat p_j\bigotimes_{j=1}^n\ket{0}^p_j=0
    \end{equation}
    Furthermore, the state $\ket{G}$ is the only common $0$-eigenvector of all the nullifiers $\hat g_i$'s if and only if $\bigotimes_{j=1}^n\ket{0_p}_j$ is the only common $0$-eigenvector of all the $\hat p_i$'s. But it's easy to see that there is only one such state. Indeed the $\hat p_i$'s are the derivative operator. And the only functions whose partial derivatives all vanish are the constant functions.  
\end{preuve}

\subsection{Proof of theorem \ref{thm: completeness gen fid}}\label{proof: completeness gen fid}

\begin{preuve}
    As shown in proposition \ref{pro: concentration g delta}, for a state verifying $O^\Delta\geq 1-\eta$, the probability $\P[\abs{\hat g_i+\delta_j}\leq x]$ that a noisy measurement of a nullifier does not yield an outcome greater than $x>0$ is bounded by
    \begin{equation}
        \P[\abs{\hat g_i+\delta_j}\leq x]\geq 1-\eta-\left(1+\frac{\delta+\Delta}{x}\right)\exp(-\frac{x^2}{(\mu+\Delta)^2}).
    \end{equation}
    When getting the outcome $x$, the state is accepted with probability $e^{-x^2/\epsilon^2}$. So for all $x>0$, we can bound from bellow,
    \begin{subequations}
        \begin{align}
            P_\text {null}&=\int_0^\infty \P[\abs{\hat g_i+\delta_j}=t]e^{-t^2/\epsilon^2}dt\\
            &\geq \int_0^x \P[\abs{\hat g_i+\delta_j}=t]e^{-t^2/\epsilon^2}dt\\
            &\geq e^{-x^2/\epsilon^2}\int_0^x \P[\abs{\hat g_i+\delta_j}=t]\\
            &=e^{-x^2/\epsilon^2} \P[\abs{\hat g_i+\delta_j}\leq x]\\
            &\geq e^{-x^2/\epsilon^2}\left(1-\eta-\left(1+\frac{\delta+\Delta}{x}\right)\exp(-\frac{x^2}{(\mu+\Delta)^2})\right)\\
            &\geq e^{-x^2/\epsilon^2}\left(1-\eta-2\exp(-\frac{x^2}{(\mu+\Delta)^2})\right) && \text{(if $x>\delta+\Delta$)}\\
            &=e^{-(\delta+\Delta)/\epsilon}\left(1-\eta-2e^{-\epsilon/(\mu+\Delta)}\right) && \text{(for $x=\sqrt{\epsilon(\delta+\Delta)}$)}.
        \end{align}
    \end{subequations}
\end{preuve}

\subsection{Proof of theorem \ref{thm: serfling}}\label{proof: operator bound}

\begin{preuve}
    To do reasoning without having to refer to the quantum formalism we introduce the random variables $Y_j^{(i)}$ whose values encode the outcome of the nullifier measurement $\hat g_i$ on register $j$, such that $Y_j^{(i)}=1$ if the measurement is rejected and $0$ otherwise. For ease of notation we also introduce $Z_j^{(i)}=1-Y_j^{(i)}$.\\

    With these notations, we can reformulate the bound we aim to show in terms of probabilities of random variables. The POVM $\hat\Pi_\text{acc}$ requires to have at least $(1-f)nN_\text{test}$ successful outcome during the initial measurement stage. We thus need,
    \begin{align}
        \sum_{i=1}^n\sum_{j\in\Pi^{(i)}} Z_j^{(i)}\geq \left(1-f\right)nN_\text{test} && \Longleftrightarrow && \sum_{i=1}^n\sum_{j\in\Pi^{(i)}} Y_j^{(i)}\leq fn N_\text{test}.
    \end{align}
    since we do a total of $nN_\text{test}$ measurements. Here the set $\Pi^{(i)}$ corresponds to the set of registers for which we have measured the nullifier $\hat g_i$. The second operator $\hat\Pi_\text{good}^N$ corresponds to checking whether there are at least $N$ registers that satisfy all nullifier tests. Since a register $j$ passes all nullifier test if and only if $\prod_{i=1}^n Z_j^{(i)}=1$, this condition can be rephrased as
    \begin{equation}
        \sum_{j\in\bar{\Pi}^{(n)}}\prod_{i=1}^n Z_j^{(i)}\geq N,
    \end{equation}
    where $\bar{\Pi}^{(i)}$ is the set of unmeasured registers after having performed the nullifier test for $\hat g_i$ (here we take $i=n$, which corresponds to the remaining unmeasured state at the end of the whole measurement step). This means that theorem \ref{thm: serfling} is equivalent to 
    \begin{equation}
        \P\left[\sum_{j\in\bar{\Pi}^{(n)}}\prod_{i=1}^n Z_j^{(i)}< N~\&~\sum_{i=1}^n\sum_{j\in\Pi^{(i)}} Y_j^{(i)}\leq fnN_\text{test}\right]\leq n \exp(-\frac{\nu^2N_\text{test}}{2}).
    \end{equation}
    To show that we will use Serfling inequality. We recall that the protocol runs as follows: Alice first chooses uniformly at random the set $\bar{\Pi}^{(1)}$ of the $N_\text{test}$ registers that will be measured by $\hat g_1$. Then she chooses $\bar{\Pi}^{(2)}$ register from the remaining one to be measured by $\hat g_2$ and so on. Serfling's bound then yields
    \begin{subequations}
        \begin{align}
            \P\left[\sum_{j\in\bar{\Pi}^{(i)}} Y_j^{(i)}< (N_\text{total}-i N_\text{test})\left(\frac{\sum_{j\in\Pi^{(i)}}Y_j^{(i)}}{N_\text{test}}+\nu\right)\right]&\geq 1-\exp[-\frac{2\nu^2(N_\text{total}-iN_\text{test})N_\text{test}^2}{[N_\text{total}-(i-1)N_\text{test}](N_\text{test}+1)}]\\
            &=1-\exp[-\frac{2\nu^2N_\text{test}}{1+N_\text{test}/(N_\text{total}-iN_\text{test})}\frac{1}{1+1/N_\text{test}}]\\
            &=1-\exp[-2\nu^2N_\text{test}\frac{1}{1+\frac{1}{\mu-i}}\frac{1}{1+\frac{1}{N_\text{test}}}]\\
            &:=q_i.
        \end{align}  
    \end{subequations}
    Denoting $E_i$ the event $\left\{\sum_{j\in\bar{\Pi}^{(i)}} Y_j^{(i)}< (N_\text{total}-i N_\text{test})\left(\frac{\sum_{j\in\Pi^{(i)}}Y_j^{(i)}}{N_\text{test}}+\nu\right)\right\}$, one wants to be able to affirm that $\P[E_1~\&~E_2~\&\cdots\&~E_n]\geq q_1\cdots q_n$. However, these events are not {\it a priori} independent. Indeed the random variables $Y_j^{(i)}$ can be correlated, and the random variable $\Pi^{(i)}$ are also correlated since Alice first chooses $\Pi^{(1)}$, then $\Pi^{(2)}$ among the unchosen registers, and so on...\\
    However, the Serfling's shows something stronger:
    \begin{equation}
        \P[E_k|\forall i,j,~Y_j^{(i)}=y_j^{(i)}]\geq q_i,
    \end{equation}
    for any set of values $y_j^{(i)}\in\{0,1\}$. Where $\P[A|B]$ denotes the conditional probability that $A$ is true knowing that $B$ is. Using this we can still bound the probability by writing
    \begin{subequations}
        \begin{align}
            \P[E_1~\&~E_2]&=\sum_{y} \P[Y=y]\P[E_1~\&~E_2|Y=y]\\
            &=\sum_{y} \sum_{\substack{\pi\subset \{1,\dots,N_\text{total}\}\\ \#\pi=N_\text{test}}} \P[Y=y]\P[\Pi^{(1)}=\pi]\P[E_1~\&~E_2|Y=y~\&~\Pi^{(1)}=\pi]\\
            &=\sum_{y} \sum_{\pi} \P[Y=y]\P[\Pi^{(1)}=\pi]\P[E_1|Y=y~\&~\Pi^{(1)}=\pi]\underbrace{\P[E_2|Y=y~\&~\Pi^{(1)}=\pi]}_{\geq q_2\text{ by Serfling}}\\
            &\geq \sum_{y} \sum_{\pi} \P[Y=y]\P[\Pi^{(1)}=\pi]\P[E_1|Y=y~\&~\Pi^{(1)}=\pi]q_2\\
            &= \sum_{y} \P[Y=y]\P[E_1|Y=y]q_2\\
            &\geq \sum_{y} \P[Y=y]q_1q_2\\
            &=q_1q_2.
        \end{align}
    \end{subequations}
    And since this reasoning can be extended to more events, we indeed have
    \begin{subequations}
        \begin{align}
            \P[E_1~\&~E_2~\&\cdots\&~E_n]&\geq q_1\cdots q_n\\ 
            &\geq q_n^n\\
            &=\left[1-\exp(-2 \nu^2N_\text{test}\frac{1}{1+1/(\mu-n)}\frac{1}{1+1/N_\text{test}})\right]^n\\
            &\geq \left[1-\exp(-\frac{\nu^2N_\text{test}}{2})\right]^n\\
            &\geq 1-n\exp(-\frac{\nu^2N_\text{test}}{2}).
        \end{align}
    \end{subequations}
    Since the event that corresponds to summing all the inequality in the events $E_i$ is more general than the event $\{E_1~\&~E_2~\&\cdots\&~E_n\}$ we deduce that, by going to the complementary event
    \begin{equation}
        \P\left[\sum_{i=1}^n\sum_{j\in\bar\Pi^{(i)}} Y_j^{(i)}\geq \sum_{i=1}^n  (N_\text{total}-i N_\text{test})\left(\frac{\sum_{j\in\Pi^{(i)}}Y_j^{(i)}}{N_\text{test}}+\nu\right)\right]\leq n\exp(-\frac{\nu^2N_\text{test}}{2}).
    \end{equation}
    Since the next event is less general than the previous one, we deduce the following inequality
    \begin{equation}
        \P\left[\sum_{i=1}^n\sum_{j\in\bar\Pi^{(i)}} Y_j^{(i)}\geq \sum_{i=1}^n  (N_\text{total}-i N_\text{test})\left(\frac{\sum_{j\in\Pi^{(i)}}Y_j^{(i)}}{N_\text{test}}+\nu\right)~\&~\sum_{i=1}^n\sum_{j\in\Pi^{(i)}} Y_j^{(i)}\leq fnN_\text{test}\right]\leq n\exp(-\frac{\nu^2N_\text{test}}{2}).
    \end{equation}
    We then get the chain of inequalities
    \begin{subequations}
        \begin{align}
            n\exp(-\frac{\nu^2N_\text{test}}{2})&\geq \P\left[\sum_{i=1}^n\sum_{j\in\bar\Pi^{(i)}} Y_j^{(i)}\geq \sum_{i=1}^n  N_\text{total}\left(\frac{\sum_{j\in\Pi^{(i)}}Y_j^{(i)}}{N_\text{test}}+\nu\right)~\&~\sum_{i=1}^n\sum_{j\in\Pi^{(i)}} Y_j^{(i)}\leq fnN_\text{test}\right]\\
            &=\P\left[\sum_{i=1}^n\sum_{j\in\bar\Pi^{(i)}} Y_j^{(i)}\geq \mu\sum_{i=1}^n  \sum_{j\in\Pi^{(i)}}Y_j^{(i)}+nN_\text{test}\mu\nu~\&~\sum_{i=1}^n\sum_{j\in\Pi^{(i)}} Y_j^{(i)}\leq fnN_\text{test}\right]\\
            &\geq\P\left[\sum_{i=1}^n\sum_{j\in\bar\Pi^{(i)}} Y_j^{(i)}\geq (f+\nu)n\mu N_\text{test}~\&~\sum_{i=1}^n\sum_{j\in\Pi^{(i)}} Y_j^{(i)}\leq fnN_\text{test}\right].
        \end{align}
    \end{subequations}
    Finally, lets consider the term $\sum_{j\in\bar{\Pi}^{(n)}}\prod_{i=1}^n Z_j^{(i)}$,
    \begin{subequations}
        \begin{align}
            \sum_{j\in\bar{\Pi}^{(n)}}\prod_{i=1}^n Z_j^{(i)}<N&\Rightarrow\sum_{j\in\bar{\Pi}^{(n)}}\prod_{i=1}^n Z_j^{(i)}\leq N\\
            &\Rightarrow \sum_{j\in\bar{\Pi}^{(n)}}\prod_{i=1}^n Z_j^{(i)}\leq nN_\text{test}(\mu/n-1-\mu f-\mu\nu)\\
            &\Rightarrow \sum_{j\in\bar{\Pi}^{(n)}} (1-Y_j^{(1)})\cdots(1-Y_j^{(n)})\leq nN_\text{test}(\mu/n-1-\mu f-\mu\nu)\\
            &\Rightarrow\underbrace{\sum_{j\in\bar{\Pi}^{(n)}} 1}_{=(\mu-n)N_\text{test}}-\underbrace{\sum_{i=1}^n\sum_{j\in\bar{\Pi}^{(n)}} Y_j^{(i)}}_{\leq\sum_{i=1}^n\sum_{j\in\bar{\Pi}^{(j)}} Y_j^{(i)}}  +\underbrace{\sum_{j\in\bar{\Pi}^{(n)}}S_j}_{\geq0}\leq nN_\text{test}(\mu/n-1-\mu f-\mu\nu)\\
            &\Rightarrow (\mu-n)N_\text{test}-\sum_{i=1}^n\sum_{j\in\bar{\Pi}^{(j)}} Y_j^{(i)}\leq nN_\text{test}(\mu/n-1-\mu f-\mu\nu)\\
            &\Rightarrow \sum_{i=1}^n\sum_{j\in\bar{\Pi}^{(j)}} Y_j^{(i)}\geq nN_\text{test}\mu(f+\nu).
        \end{align}
    \end{subequations}
    where the term $S_j$ correspond to the remaining terms in the expansion of $(1-Y_j^{(1)})\cdots(1-Y_j^{(n)})$. It is positive as for a fixed $j$, the $Y_j^{(i)}$'s take only the value $1$ or $0$. So $(1-Y_j^{(1)})\cdots(1-Y_j^{(n)})=1$ if all the $Y_j^{(i)}=0$ and $0$ otherwise. Then 
    \begin{equation}
        S_j= (1-Y_j^{(1)})\cdots(1-Y_j^{(n)})-1+ \sum_{i=1}^n S_j^{(i)}=\left\{ \begin{matrix}
            0\text{, if }Y_j^{(1)}=\cdots=Y_j^{(n)}=0\\
            Y_j^{(1)}+\cdots+Y_j^{(n)}-1\geq 0\text{, if } Y_j^{(1)}+\cdots+Y_j^{(n)}\geq 1
        \end{matrix}\right..
    \end{equation}
    So $\sum_{j\in\bar{\Pi}^{(n)}}S_j$ is indeed non negative. All of this means that \\
    
    \begin{align}
        \P\left[\sum_{j\in\bar{\Pi}^{(n)}}\prod_{i=1}^n Z_j^{(i)}< N~\&~\sum_{i=1}^n\sum_{j\in\Pi^{(i)}} Y_j^{(i)}\leq fnN_\text{test}\right]&\leq \P\left[\sum_{i=1}^n\sum_{j\in\Pi^{(i)}} Y_j^{(i)}\geq (f+\nu)n\mu N_\text{test}~\&~\sum_{i=1}^n\sum_{j\in\Pi^{(i)}} Y_j^{(i)}\leq fnN_\text{test}\right]\notag\\
        &\leq n\exp(-\frac{\nu^2N_\text{test}}{2}),
    \end{align}
    which concludes the proof.
\end{preuve}

\subsection{Proof of theorem \ref{thm: soundness}}\label{proof: soundness}
\begin{preuve}
    We start with the result of \ref{thm: serfling}
    \begin{equation}
        \Tr((\hat\Pi_\text{acc}\otimes[\1-\hat\Pi_\text{good}^N])\hat\rho)\leq n \exp(-\frac{\nu^2N_\text{test}}{1+\frac{1}{\mu-n}}),
    \end{equation}
    Using Lemma \ref{lemme: inequality operators}, we get that
    \begin{equation}
        \Tr((\hat\Pi_\text{acc}\otimes[\1-\hat\Pi_\text{good}^N])\hat\rho)\geq \Tr(\hat\Pi_\text{acc}\otimes\left[\1-\binom{N}{k}^{-1}\sum_{1\leq a_1<\cdots a_k\leq N_r}\hat \Pi_{a_1}\times\cdots\times\hat\Pi_{a_k}\right]\hat\rho).
    \end{equation}
    Next, we write
    \begin{subequations}
        \begin{align}
            \Tr(\hat\Pi_\text{acc}\otimes\sum_{1\leq a_1<\cdots a_k\leq N_r}\hat \Pi_{a_1}\times\cdots\times\hat\Pi_{a_k}\hat\rho)&=\sum_{1\leq a_1<\cdots a_k\leq N_r}\Tr(\hat\Pi_\text{acc}\otimes\hat \Pi_{a_1}\times\cdots\times\hat\Pi_{a_k}\hat\rho)\\
            &=\sum_{1\leq a_1<\cdots a_k\leq N_r}\Tr(\hat\Pi_\text{acc}\otimes\hat \Pi_1\times\cdots\times\hat\Pi_k\hat\rho),
            \intertext{Using the permutation invariance of $\hat\rho$}
            &=\binom{N_r}{k}\Tr(\hat\Pi_\text{acc}\otimes\hat \Pi_1\times\cdots\times\hat\Pi_k\hat\rho).
        \end{align}
    \end{subequations}
    So that we get
    \begin{equation}
        \Tr(\left(\hat\Pi_\text{acc}\otimes\left[\1-\frac{\binom{N_r}{k}}{\binom{N}{k}}\hat\Pi_1\times\cdots\times\hat \Pi_k\right]\right)\hat\rho)\leq n \exp(-\frac{\nu^2N_\text{test}}{1+\frac{1}{\mu-n}}),
    \end{equation}
    After that, it is simple algebra to get the various results. The lower bound for the binomial coefficient is obtained as follows
    \begin{subequations}
        \begin{align}
            \frac{\binom{N}{k}}{\binom{N_r}{k}}&=\prod_{i=0}^{k-1}\frac{N-i}{N_r-i}\\
            &\geq \left(\frac{N-k+1}{N_r-k+1}\right)^k\\
            &=\left(1-\frac{N_r-N}{N_r-k+1}\right)^k\\
            &\geq 1-k\frac{N_r-N}{N_r-k+1}\\
            &=1-\frac{knN_\text{test}\mu(f+\nu)}{(\mu-n)N_\text{test}-k+1},
        \end{align}
    \end{subequations}
    Where the first inequality is because $f(x)=\frac{N-x}{N_r-x}$ is a decreasing function. The second inequality is a consequence of $g(x)=(1-x)^k+kx-1$ being a non-negative function whenever $x\in[0,1]$. The properties of $f$ and $g$ can be obtained by simple functions analysis.
\end{preuve}

\subsection{Proof of theorem \ref{thm: fidelity bound general k systems}}\label{proof: fidelity bound k systems}
\begin{preuve}
The proof is similar to the case where we kept only one register. We start from the result of theorem \ref{thm: soundness}.
\begin{equation}
    \Tr(\hat\Pi_1\cdots\hat\Pi_k\hat\rho')\geq\left(1-\frac{knN_\text{test}\mu(f+\nu)}{(\mu-n)N_\text{test}-k+1}\right)\left(1-\frac{n\exp(-\nu^2 N_\text{test}/2)}{\Tr(\hat\Pi_\text{acc}\hat\rho)}\right).
\end{equation}
Once again, using the closure relation, we get
\begin{subequations}
    \begin{align}
        \Tr(\hat\Pi_1\cdots\hat\Pi_k\hat\rho')&=\Tr(\hat\Pi_1\cdots\hat\Pi_k\1\hat\rho')\\
        &=\int d\va s_1\cdots d\va s_k \bra{G}^{\otimes k}\hat Z(-\va s_1,\dots,-\va s_k)\hat \rho'\hat\Pi_1\cdots\hat\Pi_k\hat Z(\va s_1,\dots,\va s_k)\ket{G}^{\otimes k}\\
        &=\int d\va s_1\cdots d\va s_k \prod_{i=1}^k\prod_{j=1}^n \frac{\epsilon}{ \sqrt{\epsilon^2+\delta_j^2}}\exp(-\frac{(\va s_i)_j^2}{\epsilon^2+\delta_j^2})\bra{G}^{\otimes k}\hat Z(-\va s_1,\dots,-\va s_k)\hat \rho'\hat Z(\va s_1,\dots,\va s_k)\ket{G}^{\otimes k}\\
        &\leq \int d\va s_1\cdots d\va s_k \exp(-\frac{\norm{\va s_1}^2+\cdots+\norm{\va s_k}^2}{\Delta})\bra{G}^{\otimes k}\hat Z(-\va s_1,\dots,-\va s_k)\hat \rho'\hat Z(\va s_1,\dots,\va s_k)\ket{G}^{\otimes k}\\
        &=O_k^\Delta,
    \end{align}
\end{subequations}
whenever $\Delta^2\geq \epsilon^2+\delta_j^2$.
\end{preuve}

\subsection{Proof of proposition \ref{pro: variance not bounded}}\label{proof: variance not bounded}
\begin{preuve}
    For a state expanded as
    \begin{equation}
        \hat \rho =\int ds_1 ds_2 ds_3 ds_4 f(s_1,s_2,s_3,s_4)\hat Z(s_1,s_2)\ketbra{G}\hat Z(-s_3,-s_4),
    \end{equation}
    the density distribution for the measurement of $\hat g_1$ is 
\begin{equation}
    p(s)=\int ds_2 f(s,s_2,s,s_2),
\end{equation}
and the condition $O^\Delta\geq 1-\eta$ implies:
\begin{equation}
    1-\eta\leq O^\Delta= \int ds_1 ds_2 e^{-(s_1^2+s_2^2)/\Delta^2}f(s_1,s_2,s_1,s_2)\leq \int ds e^{-s^2/\Delta^2}p(s),
\end{equation}
So if we set $p(s)=(1-\eta)\delta(s)+\eta g(s)$ for any probability density $g(s)$, we indeed have
\begin{equation}
    \int e^{-s^2/\Delta^2}p(s)ds=(1-\eta)+\eta\int e^{-s^2/\Delta^2}g(s)\geq 1-\eta,
\end{equation}
but we have 
\begin{equation}
    \operatorname{Var}_p(s)=\eta^2 \operatorname{Var}_g(s),
\end{equation}
which can be as high as one wants, since $g$ can be any distribution. So the pure state 
\begin{equation}
    \ket{ \psi} =\int ds_1 ds_2 \sqrt{p(s_1)p(s_2)}\hat Z(s_1,s_2)\ket{G},
\end{equation}
is indeed a state satisfying $O^\Delta\geq 1-\eta$ but which has unbounded variance in the measurement of both $ \hat g_1$ and $\hat g_2$. 
\end{preuve}

\subsection{Proof of proposition \ref{pro: concentration on g}}\label{proof: concentration on g}
\begin{preuve}
    Once again, the probability density associated with the measurement of $\hat g_1$ is
    \begin{equation}
        p(s)=\int ds_2 f(s,s_2,s,s_2),
    \end{equation}
    when we write the state $\hat \rho =\int ds_1 ds_2 ds_1' ds_2' f(s_1,s_2,s_1',s_2')\hat Z(s_1,s_2)\ketbra{G}\hat Z (-s_1',-s_2')$. The condition $O^\Delta$ implies 
    \begin{equation}
        \int ds e^{-s^2/\Delta^2} p(s)\geq 1-\eta.
    \end{equation}
    We can thus write 
    \begin{subequations}
        \begin{align}
            \P[\abs{g_1}\leq x]&=\int_{-x}^x p(s) ds\\
            &\geq \int_{-x}^x e^{-s^2/\Delta^2} p(s) ds\\
            &=\int e^{-x^2/\Delta^2} p(s) ds-\int_{-\infty}^{-x}e^{-s^2/\Delta^2}p(s) ds-\int _x^\infty e^{-s^2/\Delta^2} p(s) ds\\
            &\geq 1-\eta -e^{-x^2/\Delta^2}\left(\int_{-\infty}^{-x} p(s) ds+\int _x^\infty p(s)ds\right)\\
            &\geq 1-\eta-e^{-x^2/\Delta^2}\int p(s) ds\\
            &\geq 1-\eta-e^{-x^2/\Delta^2}.
        \end{align}
    \end{subequations}
\end{preuve}

\subsection{Proof of proposition \ref{pro: concentration g delta}}\label{proof: concentration g delta}
\begin{preuve}
    We start with the proof of (\ref{eq: concentration g delta V1}). With the notation defined just before the formulation of the proposition, the hypothesis $O^\Delta\geq 1-\eta$ translates to $\int ds e^{-s^2/\Delta^2} p(s)\geq 1-\eta$. We also denote $q(s)=d\P[\hat g_i+\delta_j=s]=\frac{1}{\mu\sqrt{\pi}}\int dg p(g) e^{-(s-g)^2/\mu^2}$. We thus have:

    \begin{subequations}
        \begin{align}
            \P[\abs{\hat g_i+\delta_j}\leq x]&=\int _{-x}^x q(s)ds\\
            &\geq\int_{-x}^x e^{-\frac{x^2}{\Delta^2-\mu^2}}q(s)ds\\
            &=\int e^{-\frac{x^2}{\Delta^2-\mu^2}}q(s)ds-\int_{-\infty}^{-x}e^{-\frac{x^2}{\Delta^2-\mu^2}}q(s)ds-\int_x^\infty e^{-\frac{x^2}{\Delta^2-\mu^2}}q(s)ds\\
            &\geq \frac{1}{\mu\sqrt{\pi}}\int e^{-\frac{s^2}{\Delta^2-\mu^2}}e^{-(s-g)^2/\mu^2} p(g)dsdg-e^{-\frac{x^2}{\Delta^2-\mu^2}}\left(\int_{-\infty}^{-x} q(s)ds+\int_x^\infty q(s)ds\right)\\
            &\geq \sqrt{1-\left(\frac{\mu}{\Delta}\right)^2}\int e^{-g^2/\Delta^2}p(g)dg -e^{-\frac{x^2}{\Delta^2-\mu^2}}\int q(s) ds\\
            &\geq (1-\eta)\sqrt{1-\left(\frac{\mu}{\Delta}\right)^2}-e^{-\frac{-x^2}{\Delta^2-\mu^2}}.
        \end{align}
    \end{subequations}
    For the second inequality (\ref{eq: concentration g delta V2}), we write:
    \begin{subequations}
        \begin{align}
            \P[\abs{\hat g_i+\delta_j}\leq x ]&\geq \P[\abs{\hat g_i}\leq x/2~\&~\abs{\delta_j}\leq x/2]\\
            &=\P[\abs{\hat g_i}\leq x/2]\P[\abs{\delta_j}\leq x/2]\\
            &\geq (1-\eta -e^{-x^2/4\Delta^2})(1-\frac{2\mu}{x}e^{-x^2/4\mu^2})\\
            &\geq 1-\eta-e^{-x^2/4\Delta^2}-\frac{2\mu}{x}e^{-x^2/4\mu^2},
        \end{align}
    \end{subequations}
    using the bound on Gaussian distribution 
    \begin{subequations}
        \begin{align}
            \P[\abs{\delta_j}\leq x]&=1-2\int_x^\infty e^{-s^2/\mu^2}ds\\
            &=1-2\mu\int_{x/\mu}^\infty e^{-t^2}dt\\
            &\geq 1-\frac{\mu}{x}\int _{x/\mu}^\infty  2te^{-t^2}\\
            &=1-\frac{\mu}{x}e^{-x^2/\mu^2}.
        \end{align}
    \end{subequations}
    The last one (\ref{eq: concentration g delta V3}) is obtained by taking a generic parameter $t$ and computing
    \begin{subequations}
        \begin{align}
            \P[\abs{\hat g_i+\delta_j}\leq x ]&\geq \P[\abs{\hat g_i}\leq t~\&~\abs{\delta_j}\leq (1-t)]\\
            &=\P[\abs{\hat g_i}\leq t]\P[\abs{\delta_j}\leq (1-t)]\\
            &\geq (1-\eta -e^{-x^2t^2/\Delta^2})(1-\frac{\mu}{(1-t)x}e^{-x^2(1-t)^2/\mu^2}).
        \end{align}
    \end{subequations}
Now, if we set $t=\frac{\Delta}{\mu +\Delta}$ such that $1-t=\frac{\mu}{\Delta+\mu}$, we get

\begin{equation}
    \P[\abs{\hat g_i+\delta_j}\leq x ]\geq 1-\eta-\left(1+\frac{\mu+\Delta}{x}\right)\exp(-\frac{x^2}{(\mu+\Delta)^2}).
\end{equation}

\end{preuve}

\end{document}